\newcommand{\be}{\begin{equation}}
\newcommand{\ee}{\end{equation}}
\newcommand{\bea}{\begin{eqnarray}}
\newcommand{\eea}{\end{eqnarray}}
\newcommand{\beann}{\begin{eqnarray*}}
\newcommand{\eeann}{\end{eqnarray*}}
\newcommand{\beasn}{\begin{sneqnarray}}
\newcommand{\eeasn}{\end{sneqnarray}}
\newcommand{\ba}{\begin{array}}
\newcommand{\ea}{\end{array}}
\newcommand{\nn}{\nonumber}
\newcommand{\Appendix}[1]%
    {\renewcommand{\thesection}{Appendix~\Alph{section}:}%
     \section{#1}%
     \renewcommand{\thesection}{\Alph{section}} }

\catcode`@=11
\def\secteqno{\@addtoreset{equation}{section}%
\def\theequation{\thesection.\arabic{equation}}}
\def\endsecteqno{\def\theequation{\@ifundefined{chapter}%
{\arabic{equation}}{\thechapter.\arabic{equation}}}}
\newcounter{subequation}
\def\thesubequation{\alph{subequation}}
\def\sneqnarray{\stepcounter{equation}\let\@currentlabel=\theequation
\setcounter{subequation}{1}
\def\@eqnnum{{\rm (\theequation\thesubequation)}}
\global\@eqcnt\z@\tabskip\@centering\let\\=\@eqncr\let\@@eqncr=\@@sneqncr
$$\halign to \displaywidth\bgroup\@eqnsel\hskip\@centering
 $\displaystyle\tabskip\z@{##}$&\global\@eqcnt\@ne
 \hskip 2\arraycolsep \hfil${##}$\hfil
 &\global\@eqcnt\tw@ \hskip 2\arraycolsep
$\displaystyle\tabskip\z@{##}$\hfil
  \tabskip\@centering&\llap{##}\tabskip\z@\cr}
\def\endsneqnarray{\@@sneqncr\egroup $$\global\@ignoretrue}
\def\@@sneqncr{\let\@tempa\relax
   \ifcase\@eqcnt \def\@tempa{& & &}\or \def\@tempa{& &}
   \else \def\@tempa{&}\fi
     \@tempa \if@eqnsw\@eqnnum\stepcounter{subequation}\fi
     \global\@eqnswtrue\global\@eqcnt\z@\cr}
\def\nobiblabels{\def\@lbibitem[##1]##2{\@bibitem{##2}}}
\catcode`@=12

\def\a{\alpha}  \def\b{\beta} \def\g{\gamma} 
\def\d{\delta}  

 \def\l{\lambda}  \def\m{\mu} \def\n{\nu}

\def\s{\sigma}   
    
\def\o{\omega} \def\O{\Omega} 

\newcommand{\NP}[3]{{\it Nucl. Phys.} {\bf #1} (19#2) {#3}}

\newcommand{\PRL}[3]{{\it Phys. Rev. Lett.} {\bf #1} (19#2) {#3}}

\newcommand{\PR}[3]{{\it Phys. Rev.} {\bf #1} (19#2) {#3}}
\newcommand{\AP}[3]{{\it Ann. Phys. (N.Y.)} {\bf #1} (19#2) {#3}}
\newcommand{\AiP}[3]{{\it Adv. in Phys.} {\bf #1} (19#2) {#3}}
\newcommand{\IJMP}[3]{{\it Int. J. Mod. Phys.} {\bf #1} (19#2) {#3}}

\newcommand{\SSC}[3]{{\it Solid State Commun.} {\bf #1} (19#2) {#3}}

\newcommand{\JPCM}[3]{{\it J. Phys.: Cond. Matter} {\bf #1} (19#2) {#3}}

\def\Sv{{\bf S}}
\def\xv{{\bf x}}
\def\kv{{\bf k}}
\def\pv{{\bf p}}
\def\ev{{\hat {\bf e}}}

\def\Bv{{\bf B}}
\def\zb{{\bar z}}
\def\sv{\mbox{\boldmath $\sigma$}}
\def\SIG{\mbox{\boldmath $\Sigma$}}
\def\OME{\mbox{\boldmath $\Omega$}}
\def\Mt{{\bf M}}

\def\pa{\partial} \def\da{\dagger} 

\documentclass[12pt]{article}
\usepackage{epsfig}

\secteqno
\catcode`@=11
\long\def\@makecaption#1#2{
   \vskip 10pt
   \setbox\@tempboxa\hbox{{\small\bf #1.} \ {\small #2}}
   \ifdim \wd\@tempboxa >\hsize       
   {\small\bf #1.} \ {\small #2}\par  
   \else                              
        \hbox to\hsize{\hfil\box\@tempboxa\hfil}
   \fi}
\catcode`@=12

\begin{document}


\title{{\bf Spin Waves in Canted Phases: \\
            An Application to Doped Manganites}} 
\author{{\Large {\sl Jos\'e Mar\'{\i}a Rom\'an}}\\
        \small{\it{Department of Physics, Loomis Laboratory}}\\
        \small{\it{University of Illinois at Urbana-Champaign}}\\ 
        \small{\it{1110 W. Green St., Urbana, IL 61801-3080, USA}}\\  \\
        {\Large {\sl Joan Soto}}\\
        \small{\it{Departament d'Estructura i Constituents de la Mat\`eria 
                   and IFAE}}\\
        \small{\it{Universitat de Barcelona}}\\
        \small{\it{Diagonal, 647, E-08028 Barcelona, Catalonia, Spain.}}\\  \\
        {\it e-mails:} \small{romanfau@uiuc.edu, soto@ecm.ub.es} }
\date{\today}

\maketitle

\thispagestyle{empty}


\begin{abstract}

We present the effective lagrangian for low energy and momentum spin waves 
in canted phases at next to leading order in the derivative expansion.
The symmetry breaking pattern $SU(2)\rightarrow 1$ of the internal spin group 
and that of the crystallographic space group imply that there is one 
ferromagnetic and one antiferromagnetic spin wave. The interaction of the 
spin waves with the charge carriers is also discussed for canted,
ferromagnetic and antiferromagnetic phases. All this together allows us to
write the doping dependence of the dispersion relation parameters for doped 
manganites. We point out that the spin waves posses distinctive
characteristics  which may allow us to experimentally differentiate
canted phases from phase separation regions in doped manganites.

\end{abstract}

PACS: 75.30.Ds, 75.25.+z, 11.30.Qc, 12.39.Fe

\vfill
\vbox{
\hfill{cond-mat/9911471}\null\par
\hfill{UB-ECM-PF 98/19}\null\par}

\newpage


\section{Introduction}
\indent

Canted phases are magnetically ordered states with non-collinear
magnetisations. These configurations appear in quantum Hall double-layer 
systems \cite{Sachdev} and in the conducting regime of double 
exchange models \cite{Jonker,Anderson,DeGennes}, where the local 
magnetisations arrange in two
sublattices with magnetizations pointing to different (but not opposite) 
directions. The double exchange models are believed to provide a good
description of doped manganites \cite{Zener}, which are receiving quite a lot
of attention lately
\cite{Paco,Millis,Zou,Maezono,Golosov,Aliaga,Furukawa,Gu}. 
Doped manganites present a non-trivial interplay between their magnetic and
conducting properties \cite{Pickett,Alexandrov,Edwards,Dzero,Salamon,Tsvelik},
which leads to a rich phase diagram. The transitions between
the different phases in terms of the doping have been extensively studied 
\cite{Ramirez}.

The most studied transition is that from an antiferromagnetic insulating
phase, at zero doping, to a ferromagnetic conducting phase as the doping
grows \cite{Sheng}. It is not clear yet if the region for intermediate values 
of doping corresponds to a canted phase or to a phase separation region
\cite{DeGennes,Mn,Paco2,Kagan,Dagotto}. It is our claim that the study of the
spin waves in such materials may shed light to this question. Since the spin
waves are low energy excitations in a magnetically ordered material, they 
are sensible to the main features of the phase diagram. The 
spin waves have indeed been studied recently in connection with these materials
\cite{Matsushita,Maezono2,Loos,Gu2,Fontes,Golosov2,Pimentel,Hennion}.

The low energy and momentum dynamics of the spin waves is so much
constrained by the symmetries of the system that a model independent 
description is possible in terms of a few unknown parameters. 
For canted phases, the spontaneous symmetry breaking pattern of 
the (internal) spin symmetry is $SU(2)\rightarrow 1$, instead of 
$SU(2)\rightarrow U(1)$ like in ferromagnets or antiferromagnets. 
(Strictly speaking the symmetry breaking pattern is $SU(2)\rightarrow Z_2$, 
the center of the group. However, since we will not be concerned with
global properties, neither of the group nor of the coset manifolds, using $1$ 
instead of $Z_2$ does not modify our discussion at all). As a 
consequence of the Goldstone's theorem there will be gapless excitations in 
the spectrum (Goldstone modes) \cite{Goldstone,Guralnik}, the so called spin 
waves. A very efficient way to encode the spin waves dynamics is by using 
effective lagrangians. 

Effective lagrangians for Goldstone modes are known since the late 1960s 
\cite{Coleman}, and they have been extensively used in pion physics during the 
last decade \cite{GL}. It was suggested in ref.~\cite{Leutwyler} that they 
may also be useful in Condensed Matter systems. A detailed construction of 
the effective lagrangians for ferromagnetic and antiferromagnetic spin waves 
has already been presented in \cite{RS} (see \cite{Cliff} for a recent review
and \cite{nre,Hofmann} for non-trivial applications).

A general formalism for the spin waves in canted phases is presented in 
Section~2, where we construct an effective lagrangian at next to leading 
order. 
An intuitive separation of the spin wave field in one ferromagnetic and one 
antiferromagnetic component is also presented. Since canted phases appear 
in the conducting regime of doped manganites, the coupling of spin waves to 
charge carriers is relevant. This is discussed in Section~3, where we obtain 
an effective lagrangian for this coupling in the three different phases: 
canted, ferromagnetic and antiferromagnetic. In Section~4 we use the
previous results for the
different phases of doped manganites in order to obtain the explicit
dependence of the dispersion relation parameters on the doping, which 
is given in formulas (\ref{canted_parameters}), (\ref{F_parameters}) and
(\ref{AF_parameters}) for the canted, ferromagnetic and antiferromagnetic 
phases respectively. In Section~5 we present a plot of the doping dependence
of the velocity and the mass of the spin waves for the different phases. 
We also explain in Section~5 how our results on 
spin waves can be used to experimentally disentangle canted phases from phase 
separation regions. We summarize our conclusions in the last section. Some
properties and calculations, related to a loop integral, are relegated to the 
Appendix in order to keep our arguments clear.

In order to simplify the notation we take $\hbar = c = 1$, which leads to a
relativistic notation. Hence, we use $x = (t,\xv)$, $q=(\o,\kv)$ and 
subindices $\mu = 0,1,2,3$, where the zero stands for the time component. 
Space indices are denoted by $i = 1,2,3$.


\section{Effective Lagrangian for Canted Phases} 
\indent

In a previous paper \cite{Mn} we obtained the phase diagram for doped
manganites, where a rich set of magnetically ordered phases appeared. The
magnetically ordered configurations break spontaneously the $SU(2)$ symmetry 
of the theory (the continuum double exchange model) down to the ground state
symmetry, $U(1)$ for the ferromagnetic and antiferromagnetic configuration, 
and $1$ for the canted configuration, because of the non-collinear character 
of the latter. In this situation the lower excitations 
of the system are the spin waves, which turn out to be the Goldstone's modes 
associated to the spontaneous symmetry breaking in magnetic systems.

We have already carried out an extensive study for the ferromagnetic and 
antiferromagnetic
spin waves in crystalline solids in \cite{RS}. This formalism assumes the
existence of a gap in the excitation spectrum, which permits the construction
of an effective lagrangian for the spin waves as an expansion of local terms
suppressed by the gap. In the case that there 
are additional degrees of freedom with energy smaller than the gap, they 
should also be included in the effective lagrangian. This is the case of 
charged carriers in doped manganites, which we will discuss in section 3.
In this section, we restrict ourselves to the generalization of the formalism 
presented in \cite{RS} to the case of spin waves in canted phases.

\pagebreak


\subsection{Effective fields and symmetries}
\indent


The (internal) spin symmetry breaking pattern, $SU(2)\rightarrow 1$ for the 
non-collinear canted configurations, determines that the basic field which 
represents the Goldstone modes (spin waves) may be chosen as a 
matrix $V(x)\in SU(2)/1= SU(2)$ \cite{Coleman}. After determining the 
transformations of this field under the symmetry group of the system we 
can build an effective lagrangian from which the spin wave dynamics can be
derived. The transformations under the $SU(2)$ spin symmetry read
\be
\ba{ccl}
V(x) &  \rightarrow & g V(x) 
\ea
\quad ,\quad g\in SU(2).
\label{su2_transf}
\ee

The transformations under the crystallographic space group reduce in the 
continuum to the primitive translations and the point group. Since the local 
magnetizations in the canted phase point to two different directions depending 
on the site, both of these symmetries are broken by the ground state. This 
must be reflected in the transformation properties of $V(x)$. For definiteness,
we shall take the local magnetizations in the $1-3$ plane in the spin space, 
in such a way that the even and odd lattice magnetizations form an angle of 
$\theta /2$ and $-\theta /2$ with the $3$-axis respectively, and can be mapped 
into each other by a rotation of $\pi$ around the $3$-axis.
\bea
\Mt_1 & = & M \left(\sin(\theta/ 2), 0, \cos(\theta/ 2) \right) \nn \\
\Mt_2 & = & M \left(-\sin(\theta/ 2), 0, \cos(\theta/ 2) \right)
\label{magnetisations}
\eea

For simplicity, we shall also assume a primitive cubic lattice ($Pm\bar 3m$) 
although the ana\-ly\-sis can be carried out in a similar way for any 
crystallographic space group.
The point group $m\bar 3m$ is generated by the transformations $C_{2z}$, 
$C_{2y}$, $C_{2a}$ (a 2-fold axis in the direction $(1,1,0)$), $C_{31}^+$ 
(a 3-fold axis in the direction $(1,1,1)$) and the spatial inversion $I$. 
These transformations can be separated in two groups, on the one hand 
$\{C_{2z},C_{2y},C_{31}^+\}$, which transform points inside each sublattice, 
and on the other $\{C_{2a},I\}$, which as the primitive translations, $\tau$, 
transform points from the even sublattice to the odd one and vice-versa. Thus 
the transformation of the spin wave field $V(x)$ under this group is given by:
\be
\ba{rccl}
\xi: \{C_{2z},C_{2y},C_{31}^+\}: & V(x) & \rightarrow & g_{\xi} V(x)  \\
\xi: \{C_{2a},I\}:             & V(x) & \rightarrow & g_{\xi} V(x) R  \\
\tau: & V(x) &\rightarrow & V(x) R 
\ea
\quad ,\quad R = e^{-i\pi S^3},
\ee
where $g_{\xi}$ is the $SU(2)$ transformation associated to the point group
transformation and $R$ is a matrix which interchanges the magnetisation between
sublattices.

Notice that by combining the transformation of the field $V(x)$ under 
$\{C_{2a},I\}$ with the translations in $\{\tau C_{2a},\tau I\}$ we can 
eliminate the additional factor $R$ in those point group transformations.
Since, in addition, the factor $g_{\xi}$ can be re-absorbed by a $SU(2)$ 
transformation we only have to care about the transformations of the 
derivatives as far as the point group is concerned.

Finally, under time reversal $V(x)$ transforms as
\be
\ba{rccl}
T: & V(x) & \rightarrow & V(x) C 
\ea
\quad ,\quad C= e^{-i\pi S^2}.
\ee

We are now in a position to construct the effective lagrangian order by order 
in derivatives. In order to do that we consider the following element of the 
Lie algebra of $SU(2)$ \cite{Coleman}:
\bea
V^{\dagger}(x)i\partial_{\mu} V(x)&=& b_{\mu}^{-}(x)S_{+} + b_{\mu}^{+}(x)S_{-}
                                   + b_{\mu}^{3}(x)S^{3}.
\label{VidV}
\eea
This term, and consequently the coefficients $b_{\mu}^a(x)$, are invariant 
under the $SU(2)$
transformations (\ref{su2_transf}). Under the point group we only need to 
consider the transformation of the derivatives in $b_{\mu}^{a}(x)$, which 
correspond to the space-time indices $\mu$. The transformations under 
primitive translations are given by
\be
 \ba{rl}
 \tau: & \left\{ \ba{ccc}
      b_{\mu}^{-} & \rightarrow & -b_{\mu}^{-}   \\
      b_{\mu}^{3} & \rightarrow &  b_{\mu}^{3},
      \ea \right.
 \ea
\ee
and under time reversal
\be
   \ba{rl}
   T: &
      \left\{ \ba{ccc}
      b_{\mu}^{-} & \rightarrow & -b_{t\m}^{+}   \\
      b_{\mu}^{3} & \rightarrow & -b_{t\m}^{3},
      \ea \right.
   \ea
\label{b_time_transf}
\ee
where $t\mu$ stands for the transformation of the index $\mu$ under time 
reversal $T$.


\subsection{Relation with ferro and antiferromagnetic spin waves}
\indent

Before writing down the effective lagrangian, let us discuss a suitable 
decomposition 
of $V(x)$ which illuminates the relation between canted spin waves and the
usual ferromagnetic and antiferromagnetic ones. In the way we have chosen 
the direction of the magnetizations in each sublattice (\ref{magnetisations}), 
it is clear that the 
projection on the third direction is ferromagnetic, whereas the projection on 
the $1-2$ plane is antiferromagnetic. This suggests that we may separate the 
spin wave field into components perpendicular to the third axis and to the 
plane $1-2$ respectively. Group theory allows us to implement this easily.
Indeed, an element of the group, $V(x) \in SU(2)$, admits a unique 
decomposition in 
terms of an element of a coset, $U(x) \in SU(2)/U(1)$, and an element of the 
corresponding subgroup, $H(x) \in U(1)$, such that \mbox{$V(x) = U(x) H(x)$}, 
with
\bea 
U(x) & = & \exp\left\{ {i \over f_{\pi}}
                \left[ \pi^{-}(x)S_{+} + \pi^{+}(x)S_{-} \right] \right\}
                                          \quad \in \quad  SU(2)/U(1) \nn \\
& &   \label{UH} \\
H(x) & = & \exp\left\{{i\sqrt{2} \over f_{3}}\pi^3(x) S^3 \right\} 
                                             \quad \in  \quad U(1), \nn
\label{spin_waves}
\eea
where $S_{\pm} = S^1 \pm i S^2$ and $S^3$ are the $SU(2)$ generators,
$\pi^{\pm}(x) = \left(\pi^1(x) \pm i\pi^2(x)\right)/\sqrt{2}$ and $\pi^3(x)$
are the spin waves fields, and
$f_{\pi}$ and $f_3$ are dimensionful parameters representing the spin
stiffness. This
implies that the element of the Lie algebra in (\ref{VidV}) can be written as
\be
V^{\da}(x)i\pa_{\mu}V(x) = H^{\da}(x) (U^{\da}(x)i\pa_{\mu}U(x)) H(x) +
                                H^{\da}(x)i\pa_{\mu}H(x).  
\ee
Upon using for $U^{\dagger}(x)i\partial_{\mu} U(x)$ a similar expression
to that in (\ref{VidV}) \cite{RS} we have,
\be
U^{\dagger}(x)i\partial_{\mu} U(x) = a_{\mu}^{-}(x)S_{+} + a_{\mu}^{+}(x)S_{-}
                                   + a_{\mu}^{3}(x)S^{3}.
\label{UidU}
\ee

This decomposition translates to the coefficients $b_{\mu}^{a}(x)$ in 
(\ref{VidV}) as follows:
\bea
b_{\mu}^{-}(x) & = & e^{-i\sqrt{2}\pi^3(x)/f_3} a_{\mu}^{-}(x)  \nn \\
b_{\mu}^{+}(x) & = & e^{ i\sqrt{2}\pi^3(x)/f_3} a_{\mu}^{+}(x) \label{b:api3}\\
b_{\mu}^{3}(x) & = & a_{\mu}^{3}(x) - \sqrt{2} \pa_{\mu}\pi^3(x) / f_3.  \nn
\eea

Recall finally that the expansion of $U^{\dagger}i\partial_{\mu} U$ in spin 
wave fields reads
\be
U^{\dagger}i\partial_{\mu} U = -{1 \over f_{\pi}^2}\left[
         (f_{\pi} \pa_{\m}\pi^- + \cdots) S_+ +
         (f_{\pi} \pa_{\m}\pi^+ + \cdots) S_- +
     (i(\pi^+\pa_{\m}\pi^- - \pi^-\pa_{\m}\pi^+) + \cdots)S^3\right].
\label{expansion}
\ee


\subsection{Effective lagrangian}
\indent

In order to construct the effective lagrangian let us begin by considering 
terms with time derivatives.
It is then clear that we can build a term with a single time derivative,
\be
b_{0}^{3} = a_0^3 -{\sqrt{2} \over f_3} \pa_0 \pi^3 \sim a_0^3,
\label{1td}
\ee
which contributes to the dynamics of $\pi^{\pm}(x)$. Nevertheless 
since this term only contains a total derivative on $\pi^3(x)$ the first 
contribution to the dynamics of this field comes from
\be
b_{0}^{3} b_{0}^{3} \sim {2 \over f_3^2} \pa_0 \pi^3 \pa_0 \pi^3,
\label{2td}
\ee
where we have made explicit the quadratic term in $\pi^3(x)$.

Regarding the spatial derivatives there are no invariant terms with a single 
spatial derivative. Then the first invariant terms have two space derivatives, 
and they read
\bea
& & b_{i}^{+} b_{i}^{-} = a_{i}^{+} a_{i}^{-}  \nn \\
& & b_{i}^{+} b_{i}^{+} + b_{i}^{-} b_{i}^{-}  \label{2ts}  \\
& & b_{i}^{3} b_{i}^{3} \sim {2 \over f_3^2} \pa_i \pi^3 \pa_i \pi^3, \nn
\eea
where again we have made explicit the quadratic dependence on $\pi^3(x)$ 
in the last term.

Unlike the terms with time derivatives, the terms with spatial derivatives 
produce a leading order contributions for $\pi^{\pm}(x)$ and $\pi^3(x)$ at
the same order. Let us call it $O(p^2)$. Equations
(\ref{1td}), (\ref{2td}) and (\ref{2ts}) provide the dispersion
relations for the spin waves, which indicate how time derivatives must be
counted with respect to space derivatives. Namely, a time derivative on 
$\pi^{\pm}(x)$ must be counted as $O(p^2)$, whereas a time derivative on 
$\pi^3(x)$ must be counted as $O(p)$. This implies that the term 
$b_{0}^{3} = a_0^3 -\sqrt{2}\pa_0 \pi^3 / f_3 \sim O(p^2)+O(p)$, i.e., 
it contains terms of first and second order, which must be taken into account 
in the construction of the effective lagrangian. This is in fact a remarkable
difference with respect to the ferromagnetic and antiferromagnetic case,
where each invariant term has a unique size.

Then, putting together all the terms above, the most general effective 
lagrangian at order $O(p^2)$ we can construct, with the standard 
normalizations, reads
\bea
{\cal L}(x) & = & f_{\pi}^2 \left[ {1 \over 2} b_0^3 -B b_i^- b_i^+ 
                  - {C \over 2} (b_i^+ b_i^+ + b_i^- b_i^-) \right]
              + f_3^2 \left[ {1 \over 4} b_0^3 b_0^3 
                             - {v^2 \over 4} b_i^3 b_i^3 \right].
\label{lagp2}
\eea
If we expand it up to three fields, it reads
\bea
{\cal L}(x) & =  & \pi^- i\pa_0 \pi^+ - B \pa_i \pi^- \pa_i \pi^+ 
        -{C \over 2} (\pa_i \pi^+ \pa_i \pi^+ + \pa_i \pi^- \pa_i \pi^-) \nn \\
 & &    + {1 \over 2} \pa_0 \pi^3 \pa_0 \pi^3 
        - {v^2 \over 2} \pa_i \pi^3 \pa_i \pi^3      \label{lagsw}  \\ 
 & &    - {i\sqrt{2}C \over f_3} (\pa_i \pi^+ \pa_i \pi^+ - \pa_i \pi^- \pa_i \pi^-) \pi^3
   + {i\sqrt{2}v^2 f_3 \over 2f_{\pi}^2} (\pi^- \pa_i \pi^+ - \pi^+ \pa_i \pi^-)\pa_i \pi^3.
                                                                   \nn 
\eea

The first two lines correspond to quadratic terms in the fields, which yield 
the free propagation of the spin waves and
give the dispersion relation for each of them. Whereas these terms 
lead directly to a wave equation (Klein-Gordon type) for $\pi^3(x)$, 
as expected for an antiferromagnetic spin wave, the equation for 
$\pi^{\pm}(x)$ turns out to be non-diagonal. The off-diagonal terms are due 
to the existence in the effective lagrangian of the term 
$(b_i^+b_i^+ + b_i^-b_i^-)$. In the ferromagnetic and antiferromagnetic cases 
this term does not appear because the unbroken $U(1)$ subgroup prevents it.
In order to diagonalize the quadratic $\pi^{\pm}(x)$ terms, we perform the
following Bogolyubov transformation:
\bea
\pi^+(x) & \rightarrow & \sqrt{m^{\prime}B + {1 \over 2}} \; \pi^+(x) 
                       - \sqrt{m^{\prime}B - {1 \over 2}} \; \pi^-(x)
\quad , \quad
{1 \over 2m^{\prime}} = \sqrt{B^2 - C^2}.
\label{bogolyubov}
\eea

In terms of the new variables the lagrangian (\ref{lagsw}) reads
\bea
{\cal L}(x) & =  & \pi^- i\pa_0 \pi^+ - {1 \over 2m^{\prime}} \pa_i \pi^- \pa_i \pi^+ 
        + {1 \over 2} \pa_0 \pi^3 \pa_0 \pi^3 
        - {v^2 \over 2} \pa_i \pi^3 \pa_i \pi^3     \nn \\ 
 & &    - {i\sqrt{2}C \over f_3} (\pa_i \pi^+ \pa_i \pi^+ - \pa_i \pi^- \pa_i \pi^-) \pi^3
   + {i\sqrt{2}v^2 f_3 \over 2f_{\pi}^2} (\pi^- \pa_i \pi^+ - \pi^+ \pa_i \pi^-)\pa_i \pi^3,
\label{lagsw_diag}
\eea
which yields a Schr\"odinger equation with a mass $m^{\prime}$ for the new 
field $\pi^+(x)$. Therefore, as it was expected from the decomposition made in 
(\ref{UH})-(\ref{b:api3}) the field $\pi^+(x)$ describes one ferromagnetic spin
wave, with a quadratic dispersion relation, and $\pi^3(x)$ describes one
antiferromagnetic spin wave, with a linear dispersion relation.

This result is in agreement with previous theoretical \cite{Sachdev} and 
recent experimental \cite{Hennion} works, and in line with the general 
counting of Goldstone modes in non-relativistic systems stated in 
\cite{Nielsen} (see also \cite{Guralnik}). The general statement is that 
there exist as many real fields representing the Goldstone modes as broken 
directions in the symmetry group (three in our case, because of the 
non-collinear nature of the canted configuration). The space-time 
transformations for these fields determine if they verify a wave 
(Klein-Gordon, leading to a linear dispersion relation) 
or a Sch\"odinger (quadratic) equation of motion, with the constraint that 
in the case of a Schr\"odinger equation a complex field, and therefore two 
real ones, is necessary to represent a single physical mode (the two real 
fields behave like canonical conjugate degrees of freedom). With this 
argument in mind for the canted spin waves we can only get either
three linear branches or one linear and one quadratic branches, 
which turns out to be the correct answer in our case.

At next to leading order, $O(p^3)$, besides those terms coming from 
$b_{0}^{3} b_{0}^{3}$ already considered in (\ref{lagp2}), we find the
following terms:
\bea
& & b_0^3 b_0^3 b_0^3 \nn \\
& & b_0^3 (b_i^+ b_i^+ + b_i^- b_i^-) \nn \\
& & b_0^3 b_i^- b_i^+  \\
& & b_0^3 b_i^3 b_i^3. \nn
\eea
The term $i(b_i^- \pa_0 b_i^+ - b_i^+ \pa_0 b_i^-) \sim b_0^3 b_i^- b_i^+$ at 
this order, since at $O(p^3)$ we only have to consider time derivative acting 
on $\pi^3(x)$. However at higher orders the terms obtained from those 
invariants are different from each other.


\subsubsection{Coupling to a magnetic field}
\indent

The most important source of magnetic coupling in a spin system is the Pauli
term,  the introduction of which
in the effective theory was extensibly discussed in \cite{RS} . 
The outcome is that the Pauli term can be introduced by just 
replacing the time derivative by a covariant derivative in the following way:
\be
\pa_0 \longrightarrow D_0 \equiv \pa_0 - i \m_m \Sv \Bv.
\ee

After introducing the covariant derivative the equation (\ref{spin_waves}), 
for time derivatives, reads
\be
V^{\da}(x)iD_0 V(x) = H^{\da}(x) (U^{\da}(x)i D_0 U(x)) H(x) +
                                       H^{\da}(x)i\pa_0 H(x).  
\ee

Thus, after introducing the magnetic field the effective lagrangian is
constructed with the expressions (\ref{b:api3}) such that the
magnetic field only modifies $a_0^{\pm}(x)$ and $a_0^3(x)$, given by
\bea
\lefteqn{ U^{\da}iD_0 U \ = \ -{1 \over f_{\pi}^2} \left\{
  \left[ f_{\pi}\pa_0\pi^- -
   \m_m \left( {1 \over 2} (f_{\pi}^2 - \pi^+\pi^-)B^{\zb} +
          {1 \over 2} \pi^-\pi^- B^z + i f_{\pi} \pi^- B^3 \right) +
           \cdots \right] S_+ \right. } \nonumber \\
 & & \mbox{} + \left[ f_{\pi}\pa_0\pi^+ -
      \m_m \left( {1 \over 2} \pi^+\pi^+ B^{\zb} +
          {1 \over 2} (f_{\pi}^2 - \pi^+\pi^-)B^z -
          i f_{\pi} \pi^+ B^3 \right) + \cdots \right] S_-
                                          \label{Pauli_expansion} \\
 & & \mbox{} + \left. \biggl[ i (\pi^+\pa_0\pi^- - \pi^-\pa_0\pi^+) -
      \m_m \biggl( i f_{\pi} \pi^+ B^{\zb} -i f_{\pi}\pi^- B^z +
      (f_{\pi}^2 - 2\pi^+\pi^-) B^3 \biggr) + \cdots \biggr] S^3
                                              \right\}. \nonumber
\eea
The time derivative on $\pi^3(x)$, as well as the terms with spatial 
derivatives $b_{i}^{\pm}(x)$ and $b_i^3(x)$ remain unchanged by the presence 
of the magnetic field.

It is very easy to see that, at the lowest order, the dispersion relation of 
the antiferromagnetic branch, given by $\pi^3(x)$, is not modified by
the introduction of a small magnetic field in any direction, in particular 
in the direction of the staggered magnetization.


\subsubsection{Ferromagnetic limit}
\indent

In the ferromagnetic limit the local magnetizations are pointing in the third
direction all over the crystal. Hence, an unbroken $U(1)$ symmetry
remains, and the spin waves are represented by a field belonging to the coset
$SU(2)/U(1)$. This field can be easily obtained from the decomposition
(\ref{spin_waves}) of the canted case by taking $H(x) = 1$, or, equivalently,
\mbox{$\pi^3(x) = 0$} in (\ref{b:api3}). Hence, $V(x)$ simply reduces to 
$U(x)$. Furthermore, because of the remaining $U(1)$ symmetry terms like 
$(a_i^+a_i^+ + a_i^-a_i^-)$ are forbidden, and hence the quadratic part of the 
lagrangian does not contain off-diagonal terms. Therefore the Bogolyubov 
transformation (\ref{bogolyubov}) is not necessary anymore.

In terms of $a_{\m}^a(x)$ the effective lagrangian for the ferromagnetic spin
waves reads
\bea
{\cal L}(x) & = & f_{\pi}^2 \left[ {1 \over 2} a_0^3 
                           -{1 \over 2m^{\prime}} a_i^- a_i^+ \right].
\label{lagferrop2}
\eea
And after expanding it in terms of spin wave fields in (\ref{expansion}),
\bea
{\cal L}(x) & =  & \pi^- i\pa_0 \pi^+ 
         - {1 \over 2m^{\prime}} \pa_i \pi^- \pa_i \pi^+,
\eea
which corresponds to one spin wave with a quadratic dispersion relation.


\subsubsection{Antiferromagnetic limit}
\indent

In the antiferromagnetic limit the local magnetizations are pointing at
opposite directions in each sublattice along the first axis ($S^1$). As for the
ferromagnetic case an unbroken $U(1)$ symmetry remains, and the spin waves are
represented by an element of the coset $SU(2)/U(1)$. In order to simplify the
computation we will rotate the internal space reference frame in such a way
that the third direction, instead of the first, lies along the staggered
magnetization direction (we perform the rotation 
\mbox{$1 \rightarrow 3 \rightarrow 2$} in all the indices). With this choice 
the spin wave field is determined
from (\ref{spin_waves}) by setting $H(x) = 1$, or, equivalently,
$\pi^3(x) = 0$ in (\ref{b:api3}). $V(x)$ reduces to $U(x)$, and the remaining
$U(1)$ symmetry prevents the non-diagonal terms, 
$(a_i^+a_i^+ + a_i^-a_i^-)$, from appearing like in the ferromagnetic case.
In addition to that, now $C$ acts as the matrix which interchanges the
magnetizations between sublattices, which forbids the term with a single time
derivative, $a_0^3(x)$, to appear in the effective lagrangian.

The effective lagrangian for the antiferromagnetic spin waves is given by
\bea
{\cal L}(x) & = & f_{\pi}^2 \left[a_0^- a_0^+ - v^2 a_i^- a_i^+ \right].
\label{lagantip2}
\eea
And after expanding it in terms of spin wave fields in (\ref{expansion}),
\bea
{\cal L}(x) & =  & \pa_0 \pi^- \pa_0 \pi^+ - v^2 \pa_i \pi^- \pa_i \pi^+, 
\eea
which describes two spin waves with a linear dispersion relation.
These two branches are splitted by the introduction of a small magnetic field
in the third direction, the direction of the staggered magnetization \cite{RS}.


\section{Interaction with charge carriers}
\indent

Canted phases are known to support conductivity. Then it is important to
elucidate which kind of interaction mediates between the spin waves and the 
charge carriers. In order to address this question in a model independent way, 
we would need an effective field theory description of the latter. However, 
to our knowledge, there are no general rules on how to build such an effective 
theory, which may depend strongly on the particular material we wish to study. 
We shall then restrict ourselves to present an effective theory based on a 
particular model which successfully describes canted phases and has 
applications to doped manganites, the continuum double exchange model 
\cite{Mn}.

At first sight one may think of describing the charge carriers by an 
effective fermion field which varies slowly through the material and couples 
to the local magnetization. However, in a canted phase the local magnetization
changes abruptly from the even to the odd sublattice, which means that we shall
need two magnetization fields ${\bf M}_1(x)$ and ${\bf M}_2(x)$ in the even 
and odd sublattices, and hence a single slowly varying fermion field
is not enough to have a consistent description. We need at least two
slowly varying fermion fields $\psi_1(x)$ and $\psi_2(x)$, coupled to the 
magnetization in the even and odd sublattices respectively. 

The interaction lagrangian of the model reads
\bea
{\cal L}(x) & = & \psi^{\da}_1 (x) \left[ (1+i\epsilon)i\pa_0 + 
    {\pa^2_i \over 2m} + \m + J_H {\sv \over 2} \Mt_1 (x) \right] \psi_1 (x)  
                                                              \nn \\
& & \mbox{} + \psi^{\da}_2 (x) \left[ (1+i\epsilon)i\pa_0 + 
    {\pa^2_i \over 2m} + \m + J_H {\sv \over 2} \Mt_2 (x) \right] \psi_2 (x)  
                                        \label{lagrangian} \\
& & \mbox{} + t \left( \psi^{\da}_1 (x) \psi_2 (x) +  
           \psi^{\da}_2 (x) \psi_2 (x) \right), \nn
\eea
where $t$ corresponds to the amplitude of probability that the fermion changes 
the sublattice and $J_H$ is the Hund coupling between the fermion fields 
$\psi_1(x)$ and $\psi_2(x)$ and the magnetic moment in each sublattice 
$\Mt_1(x)$ and $\Mt_2(x)$ respectively. An estimation of our parameters is
given by $t \sim z t^l$, $J_H \sim J_H^l$ and $2m \sim 1/a^2 t^l$, where 
$a$ is the lattice spacing, $z=6$ is the coordination number and the 
superscript $l$ means the analogous lattice quantity.
In order to have conduction when $t\not= 0$ only, the chemical potential 
$\mu$ is required to lie below the lowest energy of the band for $t=0$.

The spin waves are fluctuations of the magnetically ordered ground
state, and they are included in the previous fields. We can separate the
contribution of the spin waves from that of the ground state $\Mt_1$ and 
$\Mt_2$  by writing $M_i^a(x) = R_b^a(x) M_i^b$ ($i=1,2$), such that the
matrix $R_b^a(x)$ corresponds to the spin wave fluctuation in the adjoint 
representation of $SU(2)$. Using the scalar product properties the interaction 
term can be written
\be 
{\sv \over 2} {\bf M}_{i}(x)\ =\ {\s^a \over 2} R_b^a(x) M_i^b \ = \
V(x) {\sv \over 2} V^{\da}(x){\bf M}_{i}
\qquad (i=1,2),
\ee
where the matrix $V(x)$ represents the spin waves in the fundamental 
representation of $SU(2)$.

This expression suggests the following change of variables for the fermionic 
fields
\bea
\psi_{i}(x) & \longrightarrow & V(x) \psi_{i}(x) \qquad (i=1,2).
\label{v_transf}
\eea

In terms of the new fermionic fields the lagrangian (\ref{lagrangian}) 
reads
\be
{\cal L}(x)  =  \left( \ba{cc} \psi_1^{\da}(x) & \psi_2^{\da}(x)\ea \right)
                 \left( \hat O^{gs} + \hat O^{sw} \right)
                  \left(\ba{c} \psi_1(x)  \\  \psi_2(x) \ea \right),
\label{lag_sep}
\ee
where
\be
{\hat O^{gs}} = 
\left( \ba{cc}
(1+i\epsilon)i\pa_0 + {\pa^2_i/2m} + \m + {J_H \over 2} \sv \Mt_1 
& t \\
t
& (1+i\epsilon)i\pa_0 + {\pa^2_i/2m} + \m + {J_H \over 2} \sv \Mt_2 
\ea \right)
\label{ohat}
\ee
is the contribution of the ground state, and
\be
{\hat O^{sw}} = \left( \ba{cc} 
                      {\hat{\cal O}}^{sw} & 0 \\
                       0   & {\hat{\cal O}}^{sw} \ea \right) \quad ,\quad
{\hat{\cal O}}^{sw} = (V^{\da}i\pa_0 V) 
               - {1 \over 2m} \left[ \{i\pa_i, (V^{\da}i\pa_i V) \} 
                        + (V^{\da}i\pa_i V)(V^{\da}i\pa_i V) \right],
\ee
contains the interaction with the spin waves.
The curly brackets $\{\ ,\ \}$ stand for the anti-commutator. Taking into
account the decomposition (\ref{VidV}) the operator ${\hat{\cal O}}^{sw}$ can
be expressed in terms of the fields $b_{\mu}^{a}(x)$ as follows:
\bea
{\hat{\cal O}}^{sw} & = &
       \left( b_0^- - {1 \over 2m} \{ i\pa_i , b_i^- \} \right) S_+ 
     + \left( b_0^+ - {1 \over 2m} \{ i\pa_i , b_i^+ \} \right) S_-
     + \left( b_0^3 - {1 \over 2m} \{ i\pa_i , b_i^3 \} \right) S^3   \nn \\
& &  - {1 \over 2m} \left(b_i^+ b_i^- + {1 \over 4} b_i^3 b_i^3 \right).
\label{int_sw}
\eea

A compelling expression for the coupling of the spin waves can be written by
noting that the expressions (\ref{lag_sep})-(\ref{int_sw}) are equivalent to 
the introduction of a covariant derivative,
\be
i\pa_{\mu} \ \longrightarrow \ iD_{\mu}\  =\  i\pa_{\mu} + (V^{\da}i\pa_{\mu}V)
         \ = \ i\pa_{\mu} + b_{\mu}^- S_+ + b_{\mu}^+ S_- + b_{\mu}^3 S^3,
\label{cov_der}
\ee
in (\ref{ohat}) and dropping ${\hat{\cal O}}^{sw}$.

Since the spin waves are fluctuations of long wavelength and the interaction
with the fermionic fields contains derivatives, this interaction will be small.
In this situation the problem is reduced to calculate the interaction of the
spin waves with the eigenstates of (\ref{ohat}) perturbatively. The four 
eigenstates can be obtained by considering the following change of variables:
\be
\left(\ba{c} \psi_1(x)  \\  \psi_2(x) \ea \right) =
P^{\da} \left(\ba{c} \chi_1(x)  \\  \chi_2(x) \ea \right)
\quad , \quad
P^{\da} = {1 \over \sqrt{2}} 
\left( \ba{rr}
 q + Q^{\da} + {\bar Q} &
 q - Q       - {\bar Q}  \\
 q - Q^{\da} + {\bar Q} &
-q - Q       + {\bar Q} 
\ea \right), 
\label{change}
\ee
where $q$ is an scalar parameter and $Q^{\da}$ and ${\bar Q} ={\bar Q}^{\da}$ 
are matrices in the Lie algebra of $SU(2)$ given by
\bea
q & = & {1 \over 2} \left( \sqrt{e_+ + \g + \cos{\theta \over 2} \over 2 e_+} +
          \sqrt{e_- + \g - \cos{\theta \over 2} \over 2 e_-} \right)  
\ \stackrel{\g \ll 1}{\longrightarrow} \  
{1 \over 2} \left(\cos{\theta \over 4} + \sin{\theta \over 4}\right) \nn \\
Q^{\da} & = &
{\sin{\theta \over 2} \over \sqrt{2e_- 
                    \left(e_- +\g - \cos{\theta \over 2}\right) }}\; S_+ +
{\sin{\theta \over 2} \over \sqrt{2e_+ 
                    \left(e_+ +\g + \cos{\theta \over 2}\right) }}\; S_-
\ \stackrel{\g \ll 1}{\longrightarrow} \  
\cos{\theta \over 4}\; S_+ + \sin{\theta \over 4}\; S_-  \nn\\
{\bar Q} & = & \left( \sqrt{e_+ + \g + \cos{\theta \over 2} \over 2 e_+} -
        \sqrt{e_- + \g - \cos{\theta \over 2} \over 2 e_-} \right) S^3  
\ \stackrel{\g \ll 1}{\longrightarrow} \  
\left(\cos{\theta \over 4} - \sin{\theta \over 4}\right) S^3,  \label{qs}
\eea
with
\be
e_{\pm} = \sqrt{1 + \g^2 \pm 2\g \cos{\theta \over 2}}
\qquad , \qquad
\g \equiv {2t \over \vert J_H\vert M},
\label{e_pm}
\ee
$M=\vert \Mt_1 \vert=\vert \Mt_2\vert=3/2$, and $\theta$ the angle formed by
the ground state magnetizations $\Mt_1$ and $\Mt_2$.

After the change of variables the lagrangian, written in terms of the new 
fields $\chi_1(x)$ and $\chi_2(x)$, reads 
\be
{\cal L}(x)  =  \left( \ba{cc} \chi_1^{\da}(x) & \chi_2^{\da}(x)\ea \right)
        \left[ \left( \ba{cc} {\hat L_1} &  0  \\
                                   0     & {\hat L_2} \ea  \right) +
               \left( \ba{cc} {\hat W_{11}} & {\hat W_{12}}  \\
                              {\hat W_{21}} & {\hat W_{22}} \ea \right) \right]
          \left(\ba{c} \chi_1(x)  \\  \chi_2(x) \ea \right),
\ee
where the interaction with the ground state is diagonal, and it is given by
\beasn
{\hat L_1} & = & (1+i\epsilon )i\pa_0 + {\pa_i^2 \over 2m} + \mu
  + {\vert J_H\vert M \over 2}\; \sqrt{1 + \g^2 \pm 2\g \cos{\theta \over 2}}
                                                         \label{L_1}  \\
{\hat L_2} & = & (1+i\epsilon )i\pa_0 + {\pa_i^2 \over 2m} + \mu
  - {\vert J_H\vert M \over 2}\; \sqrt{1 + \g^2 \mp 2\g \cos{\theta \over 2}}.
\eeasn
%

The interaction with the spin waves, given by 
${\hat W} = P {\hat O}^{sw} P^{\da}$, reads
\beasn
{\hat W_{11}} & = & 
 q^2 {\hat{\cal O}}^{sw} + q\{ {\hat{\cal O}}^{sw},{\bar Q} \}
            + {\bar Q} {\hat{\cal O}}^{sw} {\bar Q} 
            +  Q {\hat{\cal O}}^{sw} Q^{\da}       \\
{\hat W_{12}} & = & {\hat W_{21}^{\da}} \ = \ 
 -q[{\hat{\cal O}}^{sw}, Q] - ( {\bar Q} {\hat{\cal O}}^{sw} Q
                           +  Q {\hat{\cal O}}^{sw} {\bar Q})  \label{ws} \\
{\hat W_{22}} & = & 
 q^2 {\hat{\cal O}}^{sw} - q \{ {\hat{\cal O}}^{sw},{\bar Q} \}
            + {\bar Q} {\hat{\cal O}}^{sw} {\bar Q} 
            + Q^{\da} {\hat{\cal O}}^{sw} Q,   
\eeasn
where the square brackets $[ \ , \ ]$ stand for the commutator and 
${\hat{\cal O}}^{sw}$ is given in (\ref{int_sw}). 

In the relevant materials that we have in mind, the hopping amplitude, $t$, 
is much smaller than the Hund coupling, $J_H$, i.e. $\g \ll 1$. In this case, 
the two higher states of (\ref{ohat}), denoted by $\chi_2(x)$, lie far away 
from the two lower ones. In fact, the ratio of energies is of order $\g$. 
If we are only interested in transition energies $\sim t$ we can safely 
integrate out the states $\chi_2(x)$, obtaining the following lagrangian for
the two lowest states, $\chi_1(x)$:
\be
{\cal L}_{{\it eff}} = \chi_1^{\da}(x) 
        \left( {\hat L_1} + {\hat W_{11}} \right) \chi_1(x) -
   \chi_1^{\da}(x) {\hat W_{12}} {1 \over {\hat L_2} + {\hat W_{22}} } 
                             {\hat W_{21}} \chi_1(x).
\label{lag_int}
\ee

The second term is of order $\g$ with respect to the first one. Indeed, we
consider low
\linebreak
incoming energy and momentum with respect to the two lowest
states, namely, 
\linebreak
\mbox{${\hat L_1} \sim (1+i\epsilon )i\pa_0 + {\pa_i^2 / 2m} 
+ \mu + {\vert J_H\vert M / 2} \sim  t $}, and the spin wave interaction 
${\hat W_{ij}} \sim t$. Thus $1 / {\hat L_2} \sim \g / t$, such that the 
second term in (\ref{lag_int}) is order $\g t$, which means that the field 
$\chi_2(x)$ decouples, and the effective lagrangian reduces to
\be
{\cal L}_{{\it eff}} = \chi_1^{\da}(x) 
        \left( {\hat L_1} + {\hat W_{11}} + O(\g t) \right) \chi_1(x).
\label{lag_int2}
\ee
In order to complete the effective lagrangian we must consider the leading
order in $\g$ for ${\hat W}_{11}$ in (\ref{ws}a), which corresponds to take
the right limit in (\ref{qs}) for the parameter $q$ and the
matrices $Q^{\da}$ and $\bar Q$.

Therefore, the effective lagrangian is given by (\ref{lag_int2}), where the 
interaction terms, which come from (\ref{ws}a), (\ref{int_sw}) and (\ref{qs}), 
are given by
\bea 
{\hat W_{11}}& = & - {1 \over 2m} \left(b_i^+ b_i^- + {1 \over 4} b_i^3 b_i^3
                                               \right)  
             + {1 \over 2}\cos{\theta \over 2} 
            \left(b_0^3 - {1 \over 2m} \{ i\pa_i , b_i^3 \} \right) \nn \\
& & \mbox{} + {1 \over 2}\sin{\theta \over 2} 
\left( b_0^- - {1 \over 2m} \{ i\pa_i , b_i^- \} + 
             b_0^+ - {1 \over 2m} \{ i\pa_i , b_i^+ \} \right)
                                          \left( S_+ + S_- \right).
\label{su2_cant_sw}
\eea
Upon expanding it up to two fields, using (\ref{expansion}) and 
(\ref{b:api3}), we finally obtain
\bea
{\hat W_{11}} & = & 
-{1 \over 2m f_{\pi}^2} \left[ \pa_i\pi^- \pa_i\pi^+ 
     +{f_{\pi}^2 \over 2f_3^2} \pa_i \pi^3 \pa_i \pi^3 \right] 
                                                    \nn \\
& & \mbox{} + {1 \over 2f_{\pi}^2}\cos{\theta \over 2} \Biggl[ 
\pi^- \left( i\pa_0 \pi^+ + {1 \over 2m} \{ \pa_i,\pa_i \pi^+ \} \right)
- \pi^+ \left( i\pa_0 \pi^- + {1 \over 2m} \{ \pa_i,\pa_i \pi^- \} \right) 
                                                     \nn \\
& & \mbox{} \phantom{+ {1 \over 2f_{\pi}^2}\cos{\theta \over 2} \Biggl[ } 
+ {i\sqrt{2} f_{\pi}^2 \over f_{3}} 
\left( i\pa_0 \pi^3 + {1 \over 2m}\{ \pa_i,\pa_i \pi^3\} \right)\Biggr] 
                                                          \nn \\
& &  \mbox{} + {1 \over 2f_{\pi} f_3} \sin{\theta \over 2} \Bigg[
            if_3 \left( i\pa_0 \pi^- + i\pa_0 \pi^+ 
         + {1 \over 2m} \{ \pa_i,\pa_i \pi^- + \pa_i \pi^+ \} \right) 
                                                \label{cant_sw} \\
& & \mbox{} \phantom{+ {1 \over 2f_{\pi} f_3} \sin{\theta \over 2}\Biggl[ }
 +  \sqrt{2}\pi^3 \left( i\pa_0 \pi^- - i\pa_0 \pi^+ 
              + {1 \over 2m} \{ \pa_i,\pa_i \pi^- - \pa_i \pi^+ \} \right) 
                                                      \nn \\
& & \mbox{} \phantom{+ {1 \over 2f_{\pi} f_3} \sin{\theta \over 2} \Biggl[ }
   + {\sqrt{2} \over 2m} \pa_i\pi^3 (\pa_i \pi^- - \pa_i \pi^+)  \Biggr]
                             \left( S_+ + S_- \right).  \nn   
\eea


\subsection{Coupling to ferromagnetic spin waves}
\indent

The interaction of charge carriers with ferromagnetic spin waves can be 
considered as a limit of the canted configuration. In order to do that we must 
take the limit $\theta \rightarrow 0$ in (\ref{qs}), which yields a very simple
expression, independent of $\g$, for the parameter $q$ and the matrices
$Q^{\da}$ and $\bar Q$ which determine the change of variables (\ref{change}),
namely,
\be
q \ = \ {1 \over 2}
\qquad \qquad
Q^{\da} \ = \ S_+
\qquad \qquad
{\bar Q} \ = \ S^3.
\ee
Following the considerations in subsection~2.3.1, we also must take $H(x)=1$,
or equivalently $\pi^3(x)=0$.

Because of the remaining unbroken symmetry the $SU(2)$ transformations 
on the spin wave fields are realized by a
non-linear $U(1)_{local}$ gauge group, which allows us to write the
lagrangian in a manifestly gauge invariant way
\be
{\hat L_1} + {\hat W_{11}} = 
iD_0 +{1\over 2m}\left( D_i D_i - a^-_ia^+_i\right) +\m
+ {\vert J_{H}\vert M\over 2} + 2t S^3,
\label{su2_F_sw}
\ee
where $iD_{\m} = i\partial_{\m}+a^3_{\m}/2$. Notice that this implies that
some of the couplings are fixed by the symmetry. In fact only the coupling 
$a^-_ia^+_i$ is model dependent. This is analogous to what happens in the 
pion-nucleon lagrangian where one of the couplings is fixed by 
chiral symmetry \cite{pion}. Recall that the transformation properties under 
$U(1)_{local}$ are the following:
\bea
\chi_{i}(x) &\longrightarrow & e^{i{\varphi(x) \over 2}}\chi_{i}(x)    \nn\\
a_{\mu}^{\pm}(x) &\longrightarrow &  e^{\mp i\varphi(x)}a_{\mu}^{\pm}(x) \\
a_{\mu}^{3}(x) &\longrightarrow & a_{\mu}^{3}(x)+\partial_{\m}\varphi(x). \nn
\eea

Finally, in terms of the spin wave fields and up to two fields
${\hat W_{11}}$ reads
\bea
{\hat W_{11}} & = & 
-{1 \over 2m f_{\pi}^2} \pa_i\pi^- \pa_i\pi^+    \nn \\
& & \mbox{} + {1 \over 2f_{\pi}^2} \left[ 
\pi^- \left( i\pa_0 \pi^+ + {1 \over 2m} \{ \pa_i,\pa_i \pi^+ \} \right)
- \pi^+ \left( i\pa_0 \pi^- + {1 \over 2m} \{ \pa_i,\pa_i \pi^- \} \right)
                                                   \right],    
\label{F_sw}
\eea
which corresponds to the limit $\theta\rightarrow 0$ of the canted expression 
(\ref{cant_sw}).


\subsection{Coupling to antiferromagnetic spin waves}
\indent

In the antiferromagnetic case we have to consider two situations. The first one
corresponds to the insulating phase, where there are no charge carriers to
couple with. The second situation corresponds to the antiferromagnetic 
conducting phase. We shall describe this second situation below.

The interaction of the charge carriers with antiferromagnetic spin waves is 
given by the canted case in the limit $\theta \rightarrow \pi$. According to the
discussion in subsection~2.3.2 this limit is a little bit more involved than 
for the ferromagnetic case, since we also must rotate the reference system 
(\mbox{$1 \rightarrow 3 \rightarrow 2$}). Then the expressions for the
parameter $q$ and the matrices $Q^{\da}$ and $\bar Q$, which determine the
change of variables (\ref{change}), read
\bea
q & = & {1 \over \sqrt{2}} \sqrt{1 + {\g \over \sqrt{1 + \g^2}}} 
          \ \stackrel{\g \ll 1}{\longrightarrow} 
                              {1 \over \sqrt{2}}   \nn \\
Q^{\da} & = &  \sqrt{2 \over 1 + \g^2 +\g \sqrt{1 + \g^2}}\; S^3
        \ \stackrel{\g \ll 1}{\longrightarrow} \   \sqrt{2} S^3    \\
{\bar Q} & = & 0,                                 \nn 
\eea
where the right limit gives the leading dependence on $\g$ which we shall use
to calculate the interaction.

Similarly to the ferromagnetic case we must take $H(x)=1$, or $\pi^3(x)=0$,
and therefore the remaining unbroken symmetry determines the gauge invariance
structure for the effective lagrangian, given by
\be
{\hat L_1} + {\hat W_{11}} = 
iD_0 +{1\over 2m}\left( D_i D_i - a^-_ia^+_i\right) +\m
+ {\vert J_{H}\vert M\over 2},
\label{su2_AF_sw}
\ee
where now $iD_{\m} = i\partial_{\m}+a^3_{\m}S^3$. Notice again that this 
implies that some of the couplings are fixed by the symmetry. In fact only the
coupling $a^-_ia^+_i$ is model dependent. This is analogous to what happens 
in the ferromagnetic case discussed before and, hence, also analogous to the 
case of the pion-nucleon lagrangian \cite{pion}. Recall that the 
transformation properties under $U(1)_{local}$ are now the following:
\bea
\chi_i(x) &\longrightarrow & e^{i\varphi(x) S^3}\chi_i(x) \nn\\
a_{\mu}^{\pm}(x) &\longrightarrow &  e^{\mp i\varphi(x)}a_{\mu}^{\pm}(x)\\
a_{\mu}^{3}(x) &\longrightarrow & a_{\mu}^{3}(x)+\partial_{\m}\varphi(x). \nn
\eea

Finally, in terms of the spin wave fields and up to two fields
${\hat W_{11}}$ reads
\bea
{\hat W_{11}} & = & 
-{1 \over 2m f_{\pi}^2} \pa_i\pi^- \pa_i\pi^+    \nn \\
& & \mbox{} + {1 \over f_{\pi}^2} \left[ 
\pi^- \left( i\pa_0 \pi^+ + {1 \over 2m} \{ \pa_i,\pa_i \pi^+ \} \right)
- \pi^+ \left( i\pa_0 \pi^- + {1 \over 2m} \{ \pa_i,\pa_i \pi^- \} \right)
                                                   \right] S^3,   
\label{AF_sw}
\eea
which corresponds to the limit $\theta \rightarrow \pi$ of the canted expression 
(\ref{cant_sw}). However because of the rotation of the reference system it is
easier to obtain (\ref{AF_sw}) as a limit of (\ref{su2_cant_sw}) rather
than (\ref{cant_sw}), because the direction of the symmetry breaking has not 
been taken explicitly yet.


\section{Spin Waves in Doped Manganites}
\indent

In the section~2 we developed a general formalism which provides the 
effective lagrangian for the spin waves generated by any model in a
canted, ferromagnetic and antiferromagnetic ground state as long as the model
is invariant under $SU(2)$ transformations. In this section we are going first
to particularize this effective lagrangian to the case of the spin waves in 
doped manganites. 
Next, we will include the interaction of spin waves to charge carriers worked 
out in section 3 and calculate the doping dependence of dispersion relation
parameters.


\subsection{Spin waves from the Heisenberg hamiltonian}
\indent

In the double exchange models the interaction between the core spins in
the $t_{2g}$-bands of the manganese atom is described by an antiferromagnetic
hamiltonian. Since the value of the core spins is 
$3/2$ their motion is slow and can be approximated by classical magnetization
fields on the lattice. Furthermore, for the low energy and momentum region the
lattice fields can be further approximated by continuum fields. 
In \cite{Mn} we considered a static Heisenberg-like
interaction, which only provides the relevant contribution to the ground state 
energy. Here we shall introduce a derivative expansion of the Heisenberg 
hamiltonian, which also takes care of the low energy and momentum excitations. 
These derivative terms in the second quantization language read
\be
H = -\int{d\xv {J_{AF}a^2 \over 2z} \pa_i \Mt_1(x) \pa_i \Mt_2(x) }, 
\label{Heisenberg}
\ee
where $J_{AF} \sim z J_{AF}^l /a^3$ and the superscript $l$ represents the
lattice Heisenberg coupling. The local magnetizations for each sublattice, 
$\Mt_1(x)$ and $\Mt_2(x)$, are given as fluctuations of the ground state 
configuration, as mentioned in section~2, that we can write
$\varphi_i(x) = V(x) \varphi^{(0)}_i$,
\be
\Mt_i(x) = {\varphi^{(0)\da}_i} V^{\da}(x) \Sv V(x) \varphi^{(0)}_i =
           tr \left( V^{\da}(x) \Sv V(x) P_i \right) = R^a_b(x) M^b_i,
\ee
where $P_i$ is a projector in the direction of the ground state magnetization
in each sublattice $i = 1,2$.

This hamiltonian only generates terms with spatial derivatives in the
spin wave's effective lagrangian. In order to introduce the temporal term let
us consider it written in terms of the total, $\SIG(x)$, and staggered,
$\OME(x)$, magnetizations
\be
H = - \int{d\xv \left[ {J_{AF}a^2 \over 2z} \pa_i \SIG(x) \pa_i \SIG(x)
                   -{J_{AF}a^2 \over 8z} \pa_i \OME(x) \pa_i \OME(x) \right]},
\ee
where
\bea
\SIG(x) = {1 \over 2} \left( \Mt_1(x) + \Mt_2(x) \right)
& \qquad & \Sigma = M \cos{\theta \over 2}                     \nn\\
\OME(x) =  \Mt_1(x) - \Mt_2(x)
& \qquad & \Omega = 2M \sin{\theta \over 2}.
\eea

This corresponds to the spatial derivatives terms in the effective lagrangian
for ferromagnetic spin waves in terms of the total magnetization, $\SIG(x)$,
and for antiferromagnetic spin waves in terms of the staggered magnetization,
$\OME(x)$, \cite{Fradkin}. Following this identification we shall choose as
temporal terms those which complete these hamiltonians. Then the effective
lagrangian from the Heisenberg contribution reads
\bea
{\cal L}^{(1)}(x) & = & {1\over a^3 \Sigma^2} \int_0^1{d\l \SIG(x,\l)
              \left( \pa_0\SIG(x,\l) \times \pa_{\l}\SIG(x,\l) \right)}
            + {J_{AF}a^2 \over 2z} \pa_i \SIG(x) \pa_i \SIG(x) \nn\\
& & \mbox{} + {z \over 12 J_{AF} a^6 \O^2}
\left[ {1 \over 2} \pa_0 \OME(x) \pa_0 \OME(x) 
 - {3 J_{AF}^2 a^8 \O^2 \over 2z^2} \pa_i \OME(x) \pa_i \OME(x) \right],
\eea
where $\SIG(x,\l)$ is an extension of the total magnetization field which
verifies $\SIG(x,0) = \SIG$ and $\SIG(x,1) = \SIG(x)$. This is equivalent to
introduce an extension $\pi^a(x,\l)$ for the spin wave fields such that  
$\pi^a(x,0) = 0$ and $\pi^a(x,1) = \pi^a(x)$. A simple extension
valid for our purposes is $\pi^a(x,\l) = \l \pi^a(x)$, which allows us to
write the effective lagrangian for the canted configuration in terms of the 
spin wave representation used in section~2,
\bea
{\cal L}^{(1)}(x) & = & {2\Sigma \over a^3} \left[ {1 \over 2} b_0^3 
   + {J_{AF} a^5 \over 8z\Sigma} (8\Sigma^2 -\O^2) b_i^- b_i^+
   + {J_{AF} a^5 \O^2 \over 8z\Sigma} (b_i^+ b_i^+ + b_i^- b_i^-) \right] \nn\\
& & \mbox{}+ {z \over 6 J_{AF} a^6} \left[ {1 \over 4} b_0^3 b_0^3 
- {3J_{AF}^2 a^8 \O^2 \over 4 z^2} b_i^3 b_i^3 \right].
\label{Heisenberg_canted}
\eea

We have dropped  terms with two time derivatives
acting on $\pi^{\pm}(x)$ since they are sub-leading in the canted and 
ferromagnetic phases. However, they are not so in the antiferromagnetic 
phase and will have to be restored in order to take the 
antiferromagnetic limit. Notice that (\ref{Heisenberg_canted}) provides 
particular values for the constants $f_{\pi}^2$, $B$, $C$, $f_3^2$ and $v^2$ 
in the general formula (\ref{lagp2}).  


\subsubsection{Ferromagnetic configuration}
\indent

In the case we have the ferromagnetic configuration the limit is taken very
easily, since in this case $\Sigma \rightarrow M$ and $\O \rightarrow 0$.
Since the time evolution is already described by the term with a single time
derivative we can drop the two time derivatives term in
(\ref{Heisenberg_canted}), which yields the effective lagrangian
\bea
{\cal L}^{(1)}(x) & = & {2M \over a^3} \left[ {1 \over 2} a_0^3 
   + {J_{AF} a^5 M\over z} a_i^- a_i^+ \right].
\label{Heisenberg_F}
\eea

Notice that the mass term in (\ref{Heisenberg_F}) has the wrong sign. This is
due to the fact that (\ref{Heisenberg_F}) has been derived from an 
antiferromagnetic Heisenberg hamiltonian. Although this wrong sign apparently 
produces an instability in the 
ferromagnetic spin wave spectrum, this instability is not significant. Recall  
that the ferromagnetic phases in doped manganites are due to 
the interaction with the charge carriers. Hence any reliable estimate of the
spin wave dispersion relation parameters in the ferromagnetic phase must also 
take into account the interaction with the charge carriers. We shall do so 
later on. Such kind of (fictitious) instabilities also occur in the canted 
phases  although they are not so immediately spotted from the Lagrangian
(\ref{Heisenberg_canted}).


\subsubsection{Antiferromagnetic configuration}
\indent

In the antiferromagnetic configuration $\Sigma \rightarrow 0$ and 
$\O \rightarrow 2M$, and after performing the corresponding rotation, 
$ 1 \rightarrow 3 \rightarrow 2$, as explained in section~2, we obtain 
the following effective lagrangian:
\bea
{\cal L}^{(1)}(x) & = & {z \over 6 J_{AF} a^6} \left[  a_0^- a_0^+
                        -{12 J_{AF}^2 a^8 M^2 \over z^2}  a_i^- a_i^+ \right].
\label{Heisenberg_AF}
\eea

In this case, since the ground state configuration is supported by the
Heisenberg hamiltonian, the spin waves obtained from it are stable.


\subsection{Spin waves dispersion relations: contributions from charge 
carriers}
\indent

We have now at our disposal suitable low energy effective lagrangians 
which describe spin waves in doped manganites in the canted, ferromagnetic
and antiferromagnetic phases. They are given by the pure spin wave terms 
above together with the terms of interaction with the charge carriers 
(\ref{lag_int2}). More precisely, the effective lagrangian 
for the canted phases can be obtained from (\ref{lag_int2}),
(\ref{su2_cant_sw}) and (\ref{Heisenberg_canted}), for the ferromagnetic 
phase from (\ref{lag_int2}), (\ref{su2_F_sw}) and (\ref{Heisenberg_F}), and for
the antiferromagnetic phase from (\ref{lag_int2}), (\ref{su2_AF_sw}) and 
(\ref{Heisenberg_AF}).

In order to obtain a reliable evaluation of the parameters in the spin waves 
dispersion
relations we have to take into account the interaction with the charge carriers
in the spin waves two point Green's functions. This can be easily achieved
from a further (this time non-local) effective lagrangian which is obtained
by integrating out the charge carriers and keeping only the contributions 
up to two spin wave fields.

By integrating out the fermionic fields in the equation (\ref{lag_int2}) we 
obtain the following contributions to the effective lagrangian:
\be
S^{(2)}_{{\it eff}} = -i Tr \log({\hat L_1} + {\hat W_{11}}) =
 -i Tr \log{{\hat L_1}} -i Tr ({\hat L_1}^{-1}{\hat W_{11}}) 
 + {i \over 2} Tr ({\hat L_1}^{-1}{\hat W_{11}}{\hat L_1}^{-1}{\hat W_{11}})
 + \cdots,
\label{determinant}
\ee
where $Tr$ stands for the trace over the space-time indices as well as the
matrix indices. We have expanded the logarithm up to second order. 

The first term in (\ref{determinant}) gives rise to an effective potential
for the ground state configuration which, together with the static 
antiferromagnetic Heisenberg term, produces the rich phase diagram for doped 
manganites presented in \cite{Mn}. The following two terms in the expansion 
are responsible for the appearance of terms with at least two spin waves in 
the effective lagrangian, as can be see from (\ref{cant_sw}).

Even though in order to obtain the relevant contributions to the effective 
lagrangian it is enough to consider the interaction up to two spin waves in
(\ref{cant_sw}), interesting general characteristics will arise if, instead,
we use the $SU(2)$ invariant expression (\ref{su2_cant_sw}) for the 
interaction.
In this way we are going to obtain not only an explicitly invariant effective
lagrangian under $SU(2)$, but also the non-local structure which arises 
from the absence of gap in the fermionic spectrum of excitation.

We begin with the calculation of the second term in (\ref{determinant}), i.e. 
$S^{(2,1)}_{{\it eff}} = -i Tr ({\hat L_1}^{-1}{\hat W_{11}})$. In this
calculation a closed loop integral, representing the density of carriers, 
appears,
\be
{x \over a^3} = -i \int_{-\infty}^{\infty}{{d\o \over 2\pi} 
   {d\kv \over (2\pi)^3} tr L_1^{-1}(q) e^{i\o \eta} }.
\label{doping}
\ee
$x$ is the doping and $a^3$ the volume of the unit cell.  $tr$ represents the 
matrix trace, the space-time trace has already been taken into account in
the integration over the momentum $q=(\o, \kv)$. $ L_1^{-1}(q)$ is the
Fourier transform of the inverse of the operator $\hat L_1$ given in
(\ref{L_1}a). The convergence factor $e^{i\o \eta}$ ($\eta \rightarrow 0^+$)
is introduced to pick up the correct order of the fields in
the calculation of closed loops of one point Green's functions \cite{Fetter}.

The contribution of these terms to the effective lagrangian is given by
\be
{\cal L}^{(2,1)}(u) = {x\over 2a^3}\cos{\theta\over2} \; b_0^3 
    -{x\over 2ma^3}\left[b_i^- b_i^+ + {1 \over 4} b_i^3 b_i^3 \right],
\label{L2,1}
\ee
where we have dropped terms which contribute with a total derivative.

Whereas all the contributions in (\ref{L2,1}) are local, because the loop
integral is closed, the contribution from $S^{(2,1)}_{{\it eff}} = 
{i \over 2} Tr ({\hat L_1}^{-1}{\hat W_{11}}{\hat L_1}^{-1}{\hat W_{11}})$
is going to contain non-local terms due to the presence of the so-called
vacuum polarization tensor 
\be
\Pi^{(i,j)}_{ab}(p) = 
-i \int{{dq\over (2\pi)^4}\; (p+q)^i L^{-1}_{1a}(p+q)\; q^j L^{-1}_{1b}(q)},
\ee
where $a,b = +,-$ represent the diagonal components of the operator 
$\hat L_1$ given in (\ref{L_1}a). $i,j = 1,2,3$ represent the
spatial components of the momentum, while $i,j =0$ means the absence of the
corresponding momentum component. The properties of this tensor are displayed 
in the Appendix.

Taking into account the symmetry properties of the vacuum polarization tensor 
the contribution to the effective lagrangian reads

\bea
S^{(2,2)}_{{\it eff}} & = & -\int{du dw \int{{dp \over (2\pi)^4}\; e^{-ip(u-w)}
                                                      }} \label{S2,2}\\
& & \Biggl\{ {1\over 8} \cos^2{\theta \over 2}
       \left[ \Pi^{(0,0)}_{aa}(p)\; b_0^3(u) b_0^3(w)
            +{2\over m}\Pi^{(0,i)}_{aa}(p)\; b_0^3(u) b_i^3(w)
            +{1\over m^2}\Pi^{(i,j)}_{aa}(p)\; b_i^3(u) b_j^3(w) \right] \nn\\
& & \phantom{\Biggl\{ }
    \mbox{} + {1\over 4m^2}\sin^2{\theta \over 2}\; \Pi^{(i,j)}_{+-}(p)
        \biggl[ b_i^-(u) b_j^+(w) +  b_i^+(u) b_j^-(w)       
             + b_i^+(u) b_j^+(w) +  b_i^-(u) b_j^-(w) \biggr] \Biggr\}, \nn
\eea
where summation  convention over repeated indices has been used, and as in
the previous case terms contributing with a total derivative to the effective
lagrangian have been dropped. It is easy to see that this part 
contributes with non-local terms as long as the vacuum polarization tensor has
a non-constant behavior in the energy-momentum vector $p^{\m} =(\n,\pv)$. 
One of the most interesting terms with these characteristics is 
$b_0^3(u) b_i^3(w)$, which mixes time and spatial derivatives.

The leading contribution to the effective lagrangian is given by keeping
in (\ref{S2,2}) second order terms in derivatives (or momentum). This 
corresponds to consider the zero energy and momentum limit of the vacuum
polarization tensor, i.e. $\Pi_{ab}^{(i,j)}(0)$.
It is also convenient to choose the basis (\ref{basis}), which has its 
third component parallel to the momentum, $\pv$, in order to simplify the 
calculation. In this basis, and using the relations given in
(\ref{values}) and (\ref{sumpi33}), the action (\ref{S2,2}) reads
\bea
S^{(2,2)}_{{\it eff}} & = & -\int{du dw \int{{dp \over (2\pi)^4}\; e^{-ip(u-w)}
  \Biggl\{ {1\over 8} \cos^2{\theta \over 2}\; \Pi^{(0,0)}_{aa}(0)\; }} \nn\\
& & \phantom{-\int{du dw \int{ } } } \Biggl[b_0^3(u) b_0^3(w) 
        + 2{\n \over |\pv| } b_0^3(u) \left(e_{(3)}^i b_i^3(w)\right) 
        + \Biggl({\n \over |\pv| }\Biggr)^2 \left(e_{(3)}^i b_i^3(u)\right)
                                              \left(e_{(3)}^j b_j^3(w)\right)
                                                               \Biggr] \nn\\
& & \phantom{\Biggl\{ }
    \mbox{} -{1 \over 8}\cos^2{\theta\over 2}\; {x \over m a^3}
    \left(e_{(3)}^i b_i^3(u)\right) \left(e_{(3)}^j b_j^3(w)\right)  \nn\\
& & \phantom{\Biggl\{ }
    \mbox{} + {1\over 8} \cos^2{\theta \over 2}\; \Pi^{(1,1)}_{aa}(0)
    \biggl[  \left(e_{(1)}^i b_i^3(u)\right) \left(e_{(1)}^j b_j^3(w)\right)
          + \left(e_{(2)}^i b_i^3(u)\right) \left(e_{(2)}^j b_j^3(w)\right)
                                                          \biggr]    \\
& & \phantom{\Biggl\{ }
    \mbox{} + {1 \over 4m^2} \sin^2{\theta \over 2}\; \Pi^{(\a,\a)}_{+-}(0)
    \biggl[  \left(e_{(\a)}^i b_i^-(u)\right) \left(e_{(\a)}^j b_j^+(w)\right)
          + \left(e_{(\a)}^i b_i^+(u)\right) \left(e_{(\a)}^j b_j^-(w)\right)
                                                               \nn\\
& & \phantom{\Biggl\{ +{1 \over 4m^2}\sin^2{\theta \over 2}\;
              \Pi^{(\a,\a)}_{+-}(0) \biggl[  }  
  \mbox{} + \left(e_{(\a)}^i b_i^+(u)\right) \left(e_{(\a)}^j b_j^+(w)\right)
          + \left(e_{(\a)}^i b_i^-(u)\right) \left(e_{(\a)}^j b_j^-(w)\right)
                                                      \biggr] \Biggr\}. \nn
\eea

Hence, with the aid of (\ref{expansion}) and (\ref{b:api3}) we can expand this
expression up to two spin waves. At this order only one spin wave must be
considered in the expansion of (\ref{b:api3}), i.e. 
$b_{\m}^a \sim \pa_{\m}\pi^a$, which means that they are proportional to the
energy-momentum, and since the vectors $\ev_{(1)}$ and $\ev_{(2)}$ are 
perpendicular to the momentum, 
$e_{(\a)}^i b_i^a \sim e_{(\a)}^i \pa_i \pi^a \sim i e_{(\a)}^i p^i \pi^a = 0$,
($\a = 1,2$), they do not contribute at this order.

In addition to this, it is very interesting to notice how the terms in the
second line, which would contribute with time derivatives for $\pi^3(u)$, 
cancel at this order,
\be
{\n \over | \pv |} e_{(3)}^i b_i^3 \sim {\n p^i \over \pv^2} \pa_i \pi^3
\sim i \n \pi^3 \sim - \pa_0 \pi^3. 
\label{timetrick}
\ee
The cancellation of these terms is very important, since as it can be seen from
(\ref{values}) the tensor $\Pi^{(0,0)}_{aa}(0)$ contains an imaginary part, 
which would produce the spontaneous decay of the spin wave $\pi^3(u)$ 
into fermionic excitations.

The final result for the effective lagrangian up to two spin waves, after
using a similar procedure to (\ref{timetrick}) for spatial derivatives, turns
out to be
\bea
{\cal L}^{(2,2)}(u) & = & {J_{AF}M^2 \over 2mt f_{\pi}^2 }\; \Pi_{+-}
\sin^2 {\theta \over 2}
\left[ \pa_i\pi^-  \pa_i\pi^+ 
+ {1 \over 2} \left(\pa_i\pi^+ \pa_i\pi^+ + \pa_i\pi^- \pa_i\pi^-\right)
\right] \nn\\
& & \mbox{} + {x\over 4ma^3f_3^2}\cos^2{\theta\over 2}\; \pa_i\pi^3 \pa_i\pi^3,
\label{L2,2}
\eea
where $\Pi_{+-}$ is given in (\ref{pi_cc1}) and (\ref{pi_cc2}) for one band
canted ($CC1$) and two band canted ($CC2$) phases respectively.


\subsubsection{Ferromagnetic configuration}
\indent

The ferromagnetic limit is again very easily taken from the canted results
when \mbox{$\theta \rightarrow 0$} (equivalently we could use 
(\ref{su2_F_sw})). In addition to that, we notice from (\ref{F_sw})
that the interaction already contains at least two spin waves, thus in order
to calculate the dispersion relation we only have to consider the second term
in the expansion of the logarithm (\ref{determinant}). The effective
lagrangian at this order is
\be
{\cal L}^{(2,1)}(u) = {x\over 2a^3} \; a_0^3 -{x\over 2ma^3}a_i^- a_i^+.
\label{F_L2,1}
\ee


\subsubsection{Antiferromagnetic configuration}
\indent

In the extreme low energy and momentum limit we are interested in, there will 
be a contribution for the antiferromagnetic state in the conducting phase only.
As in the previous cases the antiferromagnetic limit of the canted expression
must be taken carefully. In this case, as in the ferromagnetic one, the
interaction already contains at least two spin waves and it is enough to
consider the second term in (\ref{determinant}). Since the
antiferromagnetic state corresponds to a 2-fold band, ${\hat L}_1$ is
degenerated, the term proportional to $S^3$ in (\ref{su2_AF_sw}) will cancel,
which prevents a term with a single time derivative from appearing in the
effective lagrangian as it should be. The final result reads
\be
{\cal L}^{(2,1)}(u) =  -{x\over 2ma^3}a_i^- a_i^+.
\label{AF_L2,1}
\ee


\subsection{Spin waves dispersion relations: final results}
\indent

Finally by summing all the contributions 
${\cal L} = {\cal L}^{(1)} + {\cal L}^{(2,1)} + {\cal L}^{(2,2)}$, given in
(\ref{Heisenberg_canted}), (\ref{L2,1}) and (\ref{L2,2}) respectively, 
we are in the position to write the effective lagrangian for the spin
waves up to second order in derivatives, and up to two spin wave fields.
After expanding the first two contributions in spin wave fields and taking
into account the expression (\ref{lagsw}) we obtain for the parameters of the
spin waves the following results:
\bea
f_{\pi}^2 & = & {2M + x \over a^3}\; y  \nn\\
B & = & {1 \over 2m}{z^{3/2} \over 15 \pi^2 A}{1 \over (2M + x)y} \left[
(1 - 3y^2) + {5A \over2}\left({6\pi^2 x \over z^{3/2}}\right)
              -  \Pi_{+-} (1- y^2) \right] \nn\\
C & = & - {1 \over 2m}{z^{3/2} \over 15 \pi^2 A}{1 \over (2M + x)y} 
                     \left[1 + \Pi_{+-} \right] (1- y^2) \nn\\
& & \label{canted_parameters}\\
f_3^2 & = & {z \over 6J_{AF} a^6} \nn\\
v^2 & = & {6 J_{AF} a^3 M^2 \over 2mz}{z^{3/2} \over 15 \pi^2 A}{1 \over M^2}
\left[2+{5A \over2}\left({6\pi^2 x \over z^{3/2}}\right)\right] (1-y^2), \nn
\eea
where $y = \cos{\theta / 2}$ is a measure of the canting angle $\theta$
and $x$ is the doping. $\Pi_{+-}$ also depends on $x$. We have used the
expression of the parameter $A$ in (\ref{xy_cc1}) and $2m \sim z/a^2t$.
In the case of the one band canted phase ($CC1$) $\Pi_{+-}$ is given by 
(\ref{pi_cc1}), and all the expressions can be written explicitly in terms of 
the doping using (\ref{xy_cc1}). In the case of the two band canted phases 
($CC2$) $\Pi_{+-}$ is given by (\ref{pi_cc2}), but it is impossible to write 
all the above expressions explicitly in terms of the doping only; we need also 
the canting angle, $y$, which depends implicitly on the doping. For a given
value of the doping we can obtain the corresponding value of $y$ by solving
the equations (\ref{xy_cc2}).

As it was described in section~2 a Bogolyubov transformation must be carried 
out in order to diagonalize the lagrangian and obtain the physical fields, 
which have a mass given by ${1 / 2m^{\prime}} = \sqrt{B^2 - C^2}$ in 
(\ref{bogolyubov}),
\be
{1\over 2m^{\prime}} = {1\over 2m}{z^{3/2}\over 15\pi^2 A}{1\over (2M + x)y}
\sqrt{2 - 4 y^2 + {5A \over2}\left({6\pi^2 x \over z^{3/2}}\right)}
\sqrt{{5A \over2}\left({6\pi^2 x \over z^{3/2}}\right) - 2y^2 
- 2 \;\Pi_{+-} (1 - y^2)}.
\label{sw_mass}
\ee

The expressions (\ref{canted_parameters}) and (\ref{sw_mass}) for the 
velocity and the mass of the spin waves are used in the plots of Fig.~1.




\subsubsection{Ferromagnetic spin waves}
\indent

The parameters for the ferromagnetic spin waves must fit (\ref{lagferrop2}),
and are obtained from the sum of (\ref{Heisenberg_F}) and (\ref{F_L2,1}).
\bea
f_{\pi}^2 & = & {2M + x \over a^3} \nn\\
{1 \over 2m^{\prime}} & = & {1 \over 2m}{z^{3/2} \over 15 \pi^2 A}
{1 \over (2M + x)} \left[
 - 2 + {5A \over2}\left({6\pi^2 x \over z^{3/2}}\right) \right],
\label{F_parameters}
\eea
which corresponds to the limit of the canted parameters
(\ref{canted_parameters}) taking into account that in the ferromagnetic limit
$1 /2m^{\prime} = B$. This is plotted in Fig.~1 in the case of a
ferromagnetic phase.

In this limit configuration it is particularly easy to see the effect of the
charge carriers in the behavior of the spin waves. Since the interaction
between the core spins is antiferromagnetic the mass derived from this
interaction is negative (producing an unstable spin wave). However, the
contribution from the fermionic sector compensates this sign and
stabilizes the spin wave. It is also worth noticing that in this limit from
the canted phase one spin wave disappears, namely, $\pi^3(u)$. Due to
the existence of the remaining $U(1)$ symmetry in the ferromagnetic ground
state this fluctuation does not modify the ground state any more.
We can see from (\ref{canted_parameters}) that the velocity of the linear 
branch goes to zero smoothly in the transition to the ferromagnetic phase. 
This answers the question asked at the end of the first paper in 
\cite{Hennion}: ``How do the spin dynamics evolve from the double-branch 
state reported here into a state with only one branch characteristic of 
a ferromagnetic metallic phase?''.


\subsubsection{Antiferromagnetic spin waves}
\indent

In this case the parameters for the effective lagrangian are obtained by
summing the contributions of (\ref{Heisenberg_AF}) and (\ref{AF_L2,1}), and
are given by
\bea
f_{\pi}^2 & = &
{z \over 6J_{AF} a^6} \nn\\
v^2 & = & {6 J_{AF} a^3 M^2\over 2mz}{z^{3/2} \over 15 \pi^2 A} {1\over M^2}
\left[2 + {5A \over2}\left({6\pi^2 x \over z^{3/2}}\right) \right],
\label{AF_parameters}
\eea
which coincides with the limit of the parameters for the antiferromagnetic
spin wave in the canted phases, $\pi^3(u)$. In the insulating phase we should
take $x=0$. This velocity is plotted in Fig.~1 for the antiferromagnetic
phase.

As in the previous case we should notice the lost of one spin wave field when
we carry out the limit towards the antiferromagnetic ground state, $\pi^1(u)$
(after the rotation of the reference system, $1 \rightarrow 3 \rightarrow 2$, 
it becomes $\pi^3(u)$). The combination of the canted phase parameters
$f_{\pi}^2 (B + C) \rightarrow 0$ with $y \rightarrow 0$. It is interesting 
to notice that the other remaining field, which used to be part of a 
ferromagnetic mode, with a quadratic dispersion relation, in the canted phase,
becomes part of the antiferromagnetic spin wave with a linear dispersion 
relation. The remaining combination of the canted phase 
$f_{\pi}^2 (B - C) \rightarrow f_3^2 v^2$ when $y \rightarrow 0$.
This answers the question at the end of the previous subsection in 
the antiferromagnetic limit.


\section{Disentangling Canted Phases from Phase Separation Regions}
\indent

Recently controversial results have appeared in the literature regarding the
existence of canted phases in doped manganites, in particular concerning
their stability against phase separation \cite{Paco,Paco2,Kagan,Dagotto}. 
We showed in \cite{Mn} that canted phases not only exist but they
are also thermodynamically stable. We presented there a phase diagram
where, in addition to stable canted phases, phase separation regions appear.

The phase diagram presents the following phases: antiferromagnetic insulating
($AFI$), antiferromagnetic conducting with two bands ($AFC2$), canted
conducting with two bands ($CC2$), canted conducting with one band ($CC1$),
ferromagnetic conducting with one band ($FC1$) and four phase separation 
regions between the $FC1$ phase and the remaining, i.e. $PS1$ ($AFI-FC1$), 
$PS2$ ($AFC2-FC1$), $PS3$ ($CC2-FC1$) and $PS4$ ($CC1-FC1$).

The question of which 
regions the system passes through when going from the antiferromagnetic 
insulating phase to the ferromagnetic conducting one upon increasing the 
doping in the actual materials could not be satisfactorily solved there, 
since the answer depends critically on the values of the
parameters of the model. For reasonable values of these parameters, various 
possibilities are allowed. We have chosen five values of the parameter $A$, 
in (\ref{xy_cc1}), for which the sequence of phases is the following:
\beann
A = 2.20 & \quad  & AFI - PS1 - FC1 \\
A = 1.75 & \quad  & AFI - AFC2 - PS2 - FC1 \\
A = 1.40 & \quad  & AFI - AFC2 - CC2 - PS3 - FC1 \\
A = 1.00 & \quad  & AFI - AFC2 - CC2 - CC1 - PS4 - FC1 \\
A = 0.80 & \quad  & AFI - AFC2 - CC2 - CC1 - FC1.
\eeann

In order to establish differences between the canted phases and the phase 
separation region we must make a guess on how these phase
separation regions look like, since our model does not describe these
non-homogeneous regions of the phase diagram. Even though these may be very
rich regions, with many different structures in them, as charge ordering, 
stripes, orbital ordering or polaronic excitations
\cite{Dagotto,Gu,Charge,Stripes,Orbital,Jahn-Teller,Mizokawa}, we shall assume
that the main structure is the coexistence of two macroscopic domains
corresponding to the phases at the border of the phase separation region, 
and that the interphase will not disturb qualitatively the properties of 
each of them. 
With this assumption in mind we have plotted in Fig.~1 the dependence
on the doping for the velocity and the mass of the spin waves for each of the
values of the parameter $A$ given above.


\begin{figure}[p]
\begin{center}
\epsfig{file=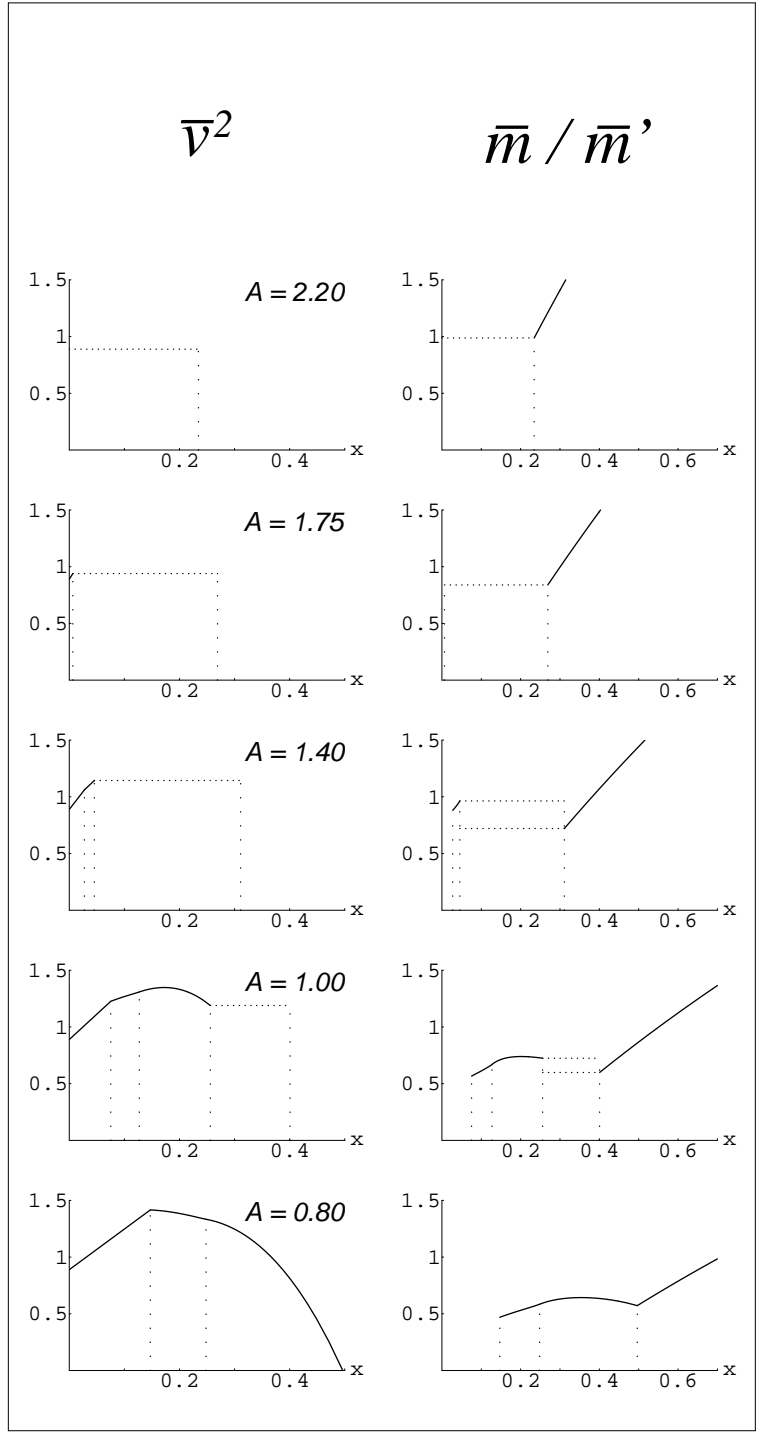,height=19cm}
\end{center}
\caption{
The dependence of the velocities and the masses with the doping for five 
different values of the parameter $A$ ($\sim t / J_{AF}$). 
${\bar v}^2 = (15 \pi^2 A) 2mz v^2 / 6z^{3/2}(J_{AF}a^3M^2)$ and 
${\bar m}/{\bar m}^{\prime} = (15 \pi^2 A) m/z^{3/2} m^{\prime}$. The 
horizontal dotted lines correspond to the phase separation regions, 
and the vertical dotted lines correspond to the phase transitions.
}
\label{fig:dispersion}
\end{figure}


The dotted lines corresponds to the values of the velocity and the mass 
in the phase separation region, for the first four values of the 
parameter $A$. They are constant values, because they
are given by the value of the corresponding phase in the border of the
phase separation region. Consider, for example, the first case, where an $AFI$
and a $FC1$ domain coexist in the phase separation region. The doping is
an extensive magnitude, and even though it reduces globally over the system,
the density of carriers remains constant in the $FC1$ domain, since this one
reduces as the doping (considered globally) decreases.

In the third and fourth cases, where canted domains coexist with the
ferromagnetic domain, we can observe two different values for the masses of
the spin waves in the phase separation region, as well as one velocity.

\medskip

Let us concentrate in the first versus the last case, i.e., the phase 
separation region with antiferromagnetic insulating and ferromagnetic 
conducting ($AFI-FC1$) domains versus the canted phases 
\cite{Paco,Paco2,Kagan,Dagotto,Biswas}. Since the differentiation 
between these two structures seems to be an experimental challenge, we 
describe below a few distinct properties of the spin waves which may help 
to differentiate between a canted phase and a phase separation region 
consisting of ferromagnetic and antiferromagnetic domains: 
\begin{enumerate}
\item[(i)] First of all, in the canted phases dispersion
relations we observe one ferromagnetic branch and one antiferromagnetic branch,
whereas in the phase separation case 
\linebreak
\mbox{($F-AF$)} we should observe one
ferromagnetic, but two antiferromagnetic branches.
\item[(ii)] Second, the antiferromagnetic branches present a further 
dramatic property:
its behavior in the presence of a magnetic field along the staggered
magnetization. Whereas in the antiferromagnetic case the two branches will be
splitted, in the canted phase the single linear branch will not be even
shifted by the presence of such a magnetic field.
\item[(iii)] Finally, we have presented in the previous section and
in Fig.~1 the different behavior of the dispersion relation parameters 
with the doping, $A = 2.20$ for the phase separation region and $A = 0.80$
for the canted phases. 
\end{enumerate}

These three characteristics, in particular, the first and the second one
which are of rather general nature (model independent), 
should allow to experimentally differentiate the regions of the 
phase diagram where ferromagnetic and antiferromagnetic phases coexist from 
those where real canted phases exist.



\section{Conclusions}
\indent

We have presented a complete study of the spin waves in canted phases.
We have exploited the spontaneous symmetry breaking pattern 
$SU(2) \rightarrow 1$ to construct an effective lagrangian for low energy and 
momentum spin 
waves in canted phases at next to leading order. For simplicity, we have 
chosen a cubic lattice, but any other lattice can be treated within the same 
formalism. The lagrangian at leading order depends on five parameters and to 
next to leading on nine. The leading lagrangian yields two spin wave modes,
one with a quadratic and one with a linear dispersion relation.
The leading effective lagrangians for ferromagnetic
and antiferromagnetic ground states were also considered as limit cases of the
canted configuration. These depend on two parameters only.

Since the canted phases appear in doped manganites, and are associated with
conducting properties of these materials
we have also presented interaction lagrangians of spin waves with charge 
carriers. Whe\-reas the lagrangian for spin waves alone is of general nature 
(model independent), the interaction with charge carriers depends on 
microscopic features of the material, in particular on the number of
conducting bands available. We have chosen a simple case (two band) which 
is inspired in a realistic model introduced in \cite{Mn} for the study of the 
phase diagram of doped manganites. Again we have derived the interaction
lagrangian for ferromagnetic and antiferromagnetic phases as the extreme cases 
of the interaction with canted spin waves. 
 
We have applied our results to the study of the spin waves in doped
manganites using the continuum double exchange model \cite{Mn}. 
We obtained the explicit dependence on the doping and the canting angle
for $f_{\pi}$, ${1 / 2m^{\prime}}$, $f_3$ and $v$, which determine the 
dispersion relations for the spin waves, and we have plotted them in Fig.~1
for several parameters.

Finally, we have proposed three ways to tell experimentally apart canted 
phases from phase separation regions (coexistence of ferromagnetic and 
antiferromagnetic phases) by looking at suitable properties of spin waves. 
The results above may also be useful for a more refined study of the phase 
diagram of doped manganites. In particular the leading quantum corrections to 
the classical spin dynamics in the low energy region are due to spin waves. It 
would be very interesting to elucidate the effect of these corrections in the 
phase diagram.


\section*{Acknowledgments}
\indent

J.M.R.~thanks Daniel S\'anchez-Portal and Eduardo Fradkin for illuminating
discussions. We thank also M.~Hennion for bringing to our attention 
ref.~\cite{Hennion}. 
J.M.R.~is supported by a Basque Government F.P.I.~postdoctoral fellowship. 
Financial support from NSF, grant no.~DMR98-17941, from CICYT (Spain), 
contract AEN98-0431 and from CIRIT (Catalonia), contract  1998SGR 00026
is also acknowledged.


\appendix


\Appendix{Vacuum Polarization Tensor}
\indent

The vacuum polarization tensor, as was defined in subsection~4.2 reads
\be
\Pi^{(i,j)}_{ab}(p) = 
-i \int{{dq\over (2\pi)^4}\; (p+q)^i L^{-1}_{1a}(p+q)\; q^j L^{-1}_{1b}(q)},
\label{tensor}
\ee
where
\be
L_{1a}^{-1}(q) = {1 \over \displaystyle \o - {\strut {\kv}^2 \over 2m} 
                                                    - \O_a + i\eta \o}
\quad , \quad
\O_{\pm} = -{\vert J_H \vert M \over 2}
              \sqrt{1 + \g^2 \pm 2\g \cos{\theta \over 2}} - \mu.
\ee
The symmetry properties of this tensor under the change of sign of the
energy and momentum are given by
\bea
\Pi^{(i,j)}_{ab}(-\nu,\pv)  & = & (-1)^{\d_{0i}+\d_{0j}} \;
                                  \Pi^{(j,i)}_{ba}(\nu,\pv) \nn\\
\Pi^{(i,j)}_{ab}(\nu,-\pv)  & = & (-1)^{\d_{0i}+\d_{0j}} \;        
                                  \Pi^{(i,j)}_{ab}(\nu,\pv)    \\
\Pi^{(i,j)}_{ab}(-\nu,-\pv) & = & \Pi^{(j,i)}_{ba}(\nu,\pv). \nn
\eea
In order to simplify the calculation of the integrals we chose a reference
system with its third component parallel to the external momentum in 
$\Pi^{(i,j)}_{ab}(\nu,\pv)$, namely,
\be
\ev_{(1)} = {({\hat \pv}{\hat \kv}){\hat \pv} - {\hat \kv} \over 
             \vert {\hat \kv} \times {\hat \pv} \vert }
\quad , \quad
\ev_{(2)} = {{\hat \kv} \times {\hat \pv} \over
             \vert {\hat \kv} \times {\hat \pv} \vert }
\quad , \quad
\ev_{(3)} = {\hat \pv},
\label{basis}
\ee
where ${\hat \kv}$ is a unit vector in the third crystallographic direction.

We will denote the components of the vectors and tensors in the new basis with
Greek indices, instead of Latin ones, such that they verify 
$w_{\a} = w_i\; e^i_{(\a)}$, and consequently,
\be
\Pi^{(\a,\b)}_{ab}(p) = \Pi^{(i,j)}_{ab}(p)\; e^i_{(\a)} e^j_{(\b)}.
\ee
Therefore the symmetry properties under the change of sign of the energy and
momentum in the new basis become
\bea
\Pi^{(\a,\b)}_{ab}(-\nu,\pv)  & = & (-1)^{\d_{0\a}+\d_{0\b}}\; 
                                  \Pi^{(\b,\a)}_{ba}(\nu,\pv) \nn\\
\Pi^{(\a,\b)}_{ab}(\nu,-\pv)  & = & (-1)^{\d_{0\a}+\d_{0\b}+\d_{1\a}+\d_{1\b}} 
                               \; \Pi^{(\a,\b)}_{ab}(\nu,\pv)    \\
\Pi^{(\a,\b)}_{ab}(-\nu,-\pv) & = & (-1)^{\d_{1\a}+\d_{1\b}}\;
                                  \Pi^{(\b,\a)}_{ba}(\nu,\pv). \nn
\eea
In this basis, after some straightforward algebra, following \cite{Fetter},
it is easy to see that all 
the dependence on the components $q^1$ and $q^2$ of the integrand in
(\ref{tensor}) comes from the explicit dependence $(p+q)^{\a}$ and $q^{\b}$,
and therefore it is verified that
\bea
\Pi^{(0,\a)}_{ab}(p)  & = & \d^{\a 3}\; \Pi^{(0,3)}_{ab}(p)  \nn\\
\Pi^{(\a,\b)}_{ab}(p) & = & \d^{\a\b}\; \Pi^{(\a,\a)}_{ab}(p),
\eea
where the repeated indices are not summed.

\medskip

Once the symmetry properties of the vacuum polarization tensor have been
considered we will address its calculation. As it was stated in the 
subsection~4.2 it is enough to calculate the limit $p^{\m} \rightarrow 0$ 
of the tensor, i.e. $\Pi^{(i,j)}_{ab}(0)$, to obtain the leading contribution
to the effective lagrangian. Since the external energy, $\n$, and momentum, 
$\pv$, injected in the fermionic loop, are related, this limit must be taken 
carefully. If they correspond to a $\pi^3(x)$ spin wave their 
relation is linear, $\n \sim \vert \pv \vert$, while the relation for 
$\pi^{\pm}(x)$ is quadratic, $\n \sim \pv^2$. The tensor components are given
by
\bea
\Pi^{(0,0)}_{aa}(0) & = & {m k_a \over 4\pi^2}
\left[-2 + x_a \log{\left| {1+x_a \over 1-x_a}\right| } 
    -i\pi \vert x_a \vert \theta(1 - \vert x_a \vert) \right]\theta(-\O_a)
\quad , \quad x_a \ = \ {m\n \over k_a \vert \pv \vert }       \nn\\
\Pi^{(0,3)}_{aa}(0) & = & 
      {m \n \over \vert \pv \vert}\; \Pi^{(0,0)}_{aa}(0)   \nn\\
\Pi^{(1,1)}_{aa}(0) & = & \Pi^{(2,2)}_{aa}(0) \ = \
  {m k_a^3 \over 12\pi^2}\; \theta(-\O_a) 
         + {1 \over 2}k_a^2(1-x_a^2)\; \Pi^{(0,0)}_{aa}(0)  \label{values} \\
\Pi^{(3,3)}_{aa}(0) & = &
   -{m k_a^3 \over 6\pi^2}\; \theta(-\O_a) 
               + \Biggl({m \n \over |\pv|} \Biggr)^2 \Pi^{(0,0)}_{aa}(0) \nn\\
\Pi^{(\a,\a)}_{+-}(0) & = & 
   -{m \over 15\pi^2}\; 
      {k_+^5 \theta(-\O_+) - k_-^5 \theta(-\O_-) \over k_+^2 - k_-^2}
\ = \  -{m J_{AF}M^2 \over t}\; \Pi_{+-}
\quad , \quad (\a = 1,2,3), \nn
\eea
where $k_a = \sqrt{-2m\O_a}$ represents the Fermi momentum in each band
$a=+,-$, and $x_a$ is the relation between the spin wave velocity and the Fermi
velocity in each band. The step functions $\theta(-\O_a)$ ensures that only 
the bands which are bellow the chemical potential contribute to the result. 
In the expressions above the summation convention was not used. An interesting
result arises from considering the summation over $\Pi^{(3,3)}_{aa}(0)$, 
namely,
\bea 
\sum_a {\Pi^{(3,3)}_{aa}(0) } & = &
-m {(2m)^{3/2} \over 6\pi^2} \sum_a (-\O_a)^{3/2} \theta(-\O_a)
+ \Biggl({m \n \over |\pv|} \Biggr)^2  \sum_a{\Pi^{(0,0)}_{aa}(0)} \nn\\
& = & -{mx \over a^3} 
+ \Biggl({m \n \over |\pv|} \Biggr)^2  \sum_a{\Pi^{(0,0)}_{aa}(0)},
\label{sumpi33}
\eea
as can be verified from an explicit calculation of the integral which gives the
doping $x$ in (\ref{doping}).

\medskip

The results obtained until now are exact (in the limit of 
$p^{\m} \rightarrow 0$). In order to obtain further analytic results for
$\Pi^{(\a,\a)}_{+-}(0)$ in (\ref{values}) we will use the value of $\O_{\pm}$
to leading order in $\g$, already obtained in \cite{Mn},
\be
\O_{\pm} = - t(y_0 \pm y),
\ee
where $y_0$ is a measure of the chemical potential.

In the $CC1$ phase we obtained in \cite{Mn} the following values for the
canting angle, $y$, and the doping, $x$:
\bea 
y & = & {5 \over 8} A (y_0 +y)^{3/2} = {5 \over 8} A 
\left({6\pi^2 x \over z^{3/2}}\right)
\qquad , \qquad
A \ = \ {z^{3/2} \over 15 \pi^2} {t \over J_{AF} a^3 M^2},
\label{xy_cc1}
\eea
and therefore
\bea
CC1: & & \Pi_{+-} \ = \ {a^3A \over z^{3/2}}\;    
{k_+^5 \over k_+^2 - k_-^2}\; \theta(-\O_+) \ = \ 
{4 \over 5} \left({6\pi^2 x \over z^{3/2}}\right)^{2/3}.
\label{pi_cc1}
\eea

The calculation is a little bit more complicated in the two band case, $CC2$,
where we obtained in \cite{Mn} for $y$ and $x$ the following expressions:
\bea
y & = & {5 \over 8} A \left[(y_0 +y)^{3/2} - (y_0 -y)^{3/2} \right] \nn\\
{5 \over 4} A (y^2 + 3y_0^2) & = & (y_0 +y)^{3/2} + (y_0 -y)^{3/2} \ = \ 
{6\pi^2 x \over z^{3/2}},
\label{xy_cc2}
\eea
which yields after eliminating the chemical potential measure, $y_0$,
\bea
CC2: & & \Pi_{+-} \ = \  {a^3A \over z^{3/2}}\;
{k_+^5 \theta(-\O_+) - k_-^5 \theta(-\O_-) \over k_+^2 - k_-^2} \ = \ 
                                              \nn \\   
& & \phantom{\Pi_{+-} } \ = \ {1 \over \sqrt{3}} {4 \over 5} 
\sqrt{{4 \over 5A} \left({6\pi^2 x \over z^{3/2}}\right)- y^2} 
+{A \over 2} \left({6\pi^2 x \over z^{3/2}}\right)
\stackrel{y \rightarrow 0}{\longrightarrow} 
{5A \over 4} \left({6\pi^2 x \over z^{3/2}}\right), 
\label{pi_cc2}
\eea
where the limit $y \rightarrow 0$ can be taken smoothly, and the previous
result is obtained by taking into account that $Ay_0^{1/2} = 8/15$ when 
$y = 0$.
               


\begin{thebibliography}{99}

\bibitem{Sachdev} S. Sachdev and T. Senthil, \AP{251}{96}{76}.
 
S. Das Sarma, S. Sachdev and L. Zheng, \PR{B58}{98}{4672};

\bibitem{Jonker} G. H. Jonker and J. H. Van Santen, {\it Physica} {\bf 50} 
(1950) 337.

\bibitem{Anderson} P. W. Anderson and H. Hasegawa, \PR{100}{55}{675}.

\bibitem{DeGennes} P. G. De Gennes, \PR{118}{60}{141}.

\bibitem{Zener} C. Zener, \PR{82}{51}{403}.

\bibitem{Paco} D. P. Arovas and F. Guinea, \PR{B58}{98}{9150}.

\bibitem{Millis} A. J. Millis, P. B. Littlewood and B. I. Shraiman, 
\PRL{75}{95}{5144}.

\bibitem{Zou} L.-J. Zou, Q.-Q. Zheng and H. Q. Lin, \PR{B56}{97}{13669}.

\bibitem{Maezono} R. Maezono, S. Ishihara and N. Nagaosa, \PR{B58}{98}{11583}.

\bibitem{Golosov} D. I. Golosov, M. R. Norman and K. Levin, \PR{B58}{98}{8617}.

\bibitem{Aliaga} H. Aliaga, R. Allub and B. Alascio, \SSC{110}{99}{525}. 

\bibitem{Furukawa} N. Furukawa, {\it Thermodynamics of the Double Exchange
Model}, cond-mat/9812066. To be published in Proc. Conference on Physics of
Manganites (July 1998 at Michigan State University).

\bibitem{Gu} R. Y. Gu, Z. D. Wang, S.-Q. Shen and D. Y. Xing, {\it Phase
diagram of an extended Kondo lattice model for manganites: the Schwinger-boson
mean-field approach}, cond-mat/9905152.

\bibitem{Pickett} W. E. Pickett and D. J. Singh \PR{B53}{96}{1146}.

\bibitem{Alexandrov} A. S. Alexandrov and A. M. Bratkovsky, 
\JPCM{11}{99}{1989}. 

\bibitem{Edwards}  D. M. Edwards, A. C. M. Green and K. Kubo, 
\JPCM{11}{99}{2791}.

\bibitem{Dzero} M. O. Dzero, L. P. Gor'kov and V. Z. Kresin, 
\SSC{112}{99}{707}.

\bibitem{Salamon} M. Jaime, P. Lin, M. B. Salamon and P. D. Han,
\PR{B58}{98}{R5901};


S. H. Chun, M. B. Salamon, Y. Lyanda-Geller, P. M. Goldbart and P. D. Han,
{\it Phys. Rev. Lett.} {\bf 84} (2000) 757; 

M. Jaime and M. B. Salamon, {\it Electronic Transport in La-Ca Manganites}
cond-mat/9902284. To be published in the proceedings of the Workshop
on Physics of Manganites, MSU, July 1998

\bibitem{Tsvelik} O. Zachar, A. M. Tsvelik, {\it One-dimensional electron gas 
interacting with a Heisenberg spin-1/2 chain}, cond-mat/9909296.

\bibitem{Ramirez} A. P. Ramirez, \JPCM{9}{97}{8171}.

J. M. D. Coey, M. Viret and S. von Molnar, \AiP{40}{99}{167}. 

\bibitem{Sheng} L. Sheng, D. N. Sheng, C. S. Ting, 
\PR{B59}{99}{13550};

P. Raychaudhuri, C. Mitra, A. Paramekanti, R. Pinto, A. K. Nigam and 
S. K. Dhar, \JPCM{10}{98}{L191}.

\bibitem{Mn} J. M. Rom\'an and J. Soto, \PR{B59}{99}{11418}.

\bibitem{Paco2} D. P. Arovas, G. G\'omez-Santos and F. Guinea, 
\PR{B59}{99}{13569}.

\bibitem{Kagan} M. Yu. Kagan, D. I. Khomskii and M. V. Mostovoy, 
{\it European Physical Journal} {\bf B12} (1999) 217.

\bibitem{Dagotto} E. Dagotto, S. Yunoki and A. Moreo, {\it Phase separation in 
models for manganites: theoretical aspects and comparation with experiments}, 
cond-mat/9809380. To appear in proceedings of the workshop
``Physics of Manganites'', Michigan State University, July 26--29, 1998, eds. 
by T. A. Kaplan and S. D. Mahanti, Plenum Publishing Corporation;

S. Yunoki, A. Moreo and E. Dagotto, \PRL{81}{98}{5612}; 

A. Moreo, S. Yunoki and E. Dagotto, {\it The Phase Separation Scenario for 
Manganese Oxides}, cond-mat/9901057. To be published in {\it Science}.

\bibitem{Matsushita} Y. Matsushita, R. Shiina and C. Ishii, \SSC{97}{96}{71}.

\bibitem{Maezono2} R. Maezono and N. Nagaosa, 
{\it Phys. Rev.} {\bf B61} (2000) 1189.

\bibitem{Loos} J. Loos and H. Fehske, 
{\it Physica} {\bf B259-261} (1999) 801.

\bibitem{Gu2} R. Y. Gu, Shun-Qing Shen, Z. D. Wang and D. Y. Xing,
{\it Linear spin and orbital wave theory for undoped manganites},
cond-mat/9908464.

\bibitem{Fontes} M. B. Fontes, J. C. Trochez, B. Giordanengo, S. L. Bud'ko,
D. R. Sanchez, E. M. Baggio-Saitovitch and M. A. Continentino, 
\PR{B60}{99}{6781}.

\bibitem{Golosov2} D. I. Golosov, {\it Spin Wave Theory of Double Exchange
Magnets}, cond-mat/9909213.

\bibitem{Pimentel} I. R. Pimentel, F. Carvalho, L. M. Martelo and R. Orbach, 
\PR{B60}{99}{12329}.

\bibitem{Hennion} F. Moussa, M. Hennion, G. Biotteau, 
J. Rodr\'{\i}guez-Carvajal, L. Pinsard and A. Revcolevschi,
\PR{B60}{99}{12299};

M. Hennion, F. Moussa, G. Biotteau, J. Rodr\'{\i}guez-Carvajal, L. Pinsard and 
A. Revcolevschi, {\it Phys. Rev.} {\bf B61} (2000) 9513.

\bibitem{Goldstone} J. Goldstone, {\sl Nuovo Cim.} {\bf 19} (1961) 145.

\bibitem{Guralnik} G. S. Guralnik, C. R. Hagen and T. W. B. Kibble,
{\sl Advances in particle physics}, Vol.2, p.567, ed. R. L. Cool and
R. E. Marshak (Willey, New York, 1968).

\bibitem{Coleman} S. Coleman, J. Wess and B. Zumino, \PR{177}{69}{2239};

C. Callan, S. Coleman, J. Wess and B. Zumino, \PR{177}{69}{2247}.

\bibitem{GL} J. Gasser and H. Leutwyler, \AP{158}{84}{142}.

\bibitem{Leutwyler} H. Leutwyler, \PR{D49}{94}{3033}. 

\bibitem{RS} J. M. Rom\'an and J. Soto, \IJMP{B13}{99}{755}.

\bibitem{Cliff} C. P. Burgess, {\it Goldstone and Pseudo-Goldstone bosons in 
Nuclear, Particle and Condensed Matter Physics}, hep-th/9808176. 

\bibitem{nre} J. M. Rom\'an and J. Soto, \AP{273}{99}{37}.

\bibitem{Hofmann} C. P. Hofmann, {\it Thermodynamic Behavior of the $O(3)$
Ferromagnet}, Universit\"at Bern, preprint no. BUTP-97/31; 

C. P. Hofmann, \PR{B60}{99}{388};

C. P. Hofmann, \PR{B60}{99}{406}.

\bibitem{Nielsen} H. B. Nielsen and S. Chada, \NP{B105}{76}{445}.

\bibitem{pion} J.F. Donoghue, E. Golowich and B.R. Holstein, 
{\it Dynamics of the Standard Model} (Section {\bf IV-7}), 
(Cambridge University Press, 1992).

\bibitem{Fradkin} E. Fradkin, {\it Field Theories of Condensed Matter
Systems} (Addison-Wesley, Reading, MA, 1990);

S. Chakravarty, B. I. Halperin and D. R. Nelson,
\PR{B39}{89}{2344}.

\bibitem{Fetter} A. L. Fetter and J. D. Walecka, {\it Quantum Theory of
Many-Particle Systems}, (McGraw-Hill, New York, 1971).

\bibitem{Charge} S. Yunoki, T. Hotta and E. Dagotto, 
{\it Ferromagnetic, A-type, and Charge-Ordered CE-type States in Doped 
Manganites using Jahn-Teller Phonons}, cond-mat/9909254.

\bibitem{Stripes} T. Mutou and H. Kontani, \PRL{83}{99}{3685}.

\bibitem{Orbital} L. Sheng and C. S. Ting, {\it Modified Double Exchange Model 
with Novel Spin and Orbital Coupling: Phase Diagram of The Manganites}, 
cond-mat/9812374;

J. van den Brink and D. Khomskii, \PRL{82}{99}{1016}.

\bibitem{Jahn-Teller} P. Benedetti and R. Zeyher, \PR{B59}{99}{9923};

T. Hotta, S. Yunoki and E. Dagotto, {\it Cooperative Jahn-Teller Effect on the 
Magnetic Structure of Manganese Oxides}, cond-mat/9907430. 
To appear in proceedings of the
conference "Science and Technology of Magnetic Oxides '99", La Jolla, 
July 5--7, 1999.

\bibitem{Mizokawa} T. Mizokawa, D. I. Khomskii and G. A. Sawatzky, 
\PR{B60}{99}{7309}.

\bibitem{Biswas} A. Biswas, M. Rajeswari, R. C. Srivastava,  Y. H. Li, 
T. Venkatesan, R. L. Greene and A. J. Millis, 
{\it Phys. Rev.} {\bf B61} (2000) 9665.

\end{thebibliography}
\end{document}